\documentclass[12pt,noshowpacs,nofootinbib,notitlepage,amsmath,superscriptaddress]{revtex4-1}

\usepackage{slashed}
\usepackage[utf8]{inputenc}
\usepackage[english]{babel}    
\usepackage{tocvsec2}
\usepackage{amsthm} 
\pdfoutput=1
\allowdisplaybreaks
\usepackage{graphicx,color}
\usepackage[colorlinks=true,citecolor=blue,linkcolor=blue,urlcolor=blue]{hyperref}
\usepackage{comment}
\usepackage{appendix}

 \usepackage{graphicx, psfrag} 

\usepackage{footmisc}
\usepackage{bm}
\usepackage{dcolumn}
\usepackage{gensymb}
 \usepackage{float}
\restylefloat{table}
\usepackage{multirow}
 \usepackage{hyperref}
\hypersetup{
    colorlinks,
    citecolor=black,
    filecolor=black,
    linkcolor=blue,
    urlcolor=blue
}
\usepackage{epsfig}
\def\be{\begin{equation}}
\def\ee{\end{equation}}
\def\bea{\begin{eqnarray}}
\def\eea{\end{eqnarray}}
\def\ba{\begin{equation}\begin{aligned}}

\usepackage{color}
\usepackage[dvipsnames]{xcolor}

\definecolor{darkblue}{rgb}{0.0, 0.0, 0.55}
\definecolor{grey}{rgb}{0.57, 0.64, 0.69}
\definecolor{lightbrown}{rgb}{0.71, 0.4, 0.11}

\begin{document}
\title{Stellar Structure and Stability of Charged Interacting Quark Stars and Their Scaling Behaviour}
\author{Chen Zhang}
\email{zhangvchen@gmail.com}
\affiliation{Department of Physics and Astronomy, University of Waterloo, Waterloo, Ontario, N2L 3G1, Canada}
\author{Michael Gammon}
\email{gammon.michael10@gmail.com}
\affiliation{Department of Physics and Astronomy, University of Waterloo, Waterloo, Ontario, N2L 3G1, Canada}
\author{Robert B. Mann}
\email{rbmann@uwaterloo.ca}
\affiliation{Department of Physics and Astronomy, University of Waterloo, Waterloo, Ontario, N2L 3G1, Canada}


\begin{abstract}
We explore the stellar structure and radial stability of charged quark stars composed of interacting quark matter in three classes of commonly used charge models. We adopt a general parametrization of interacting quark matter equation of state that includes the corrections from perturbative QCD, color superconductivity, and the strange quark mass into one parameter $\lambda$, or one dimensionless parameter $\bar{\lambda}=\lambda^2/(4B_{\rm eff})$ after being rescaled with the effective bag constant $B_{\rm eff}$.  We find that increasing charge tends to increase the mass and radius profiles, and enlarges the separation size in mass  between the maximum mass point and the point where zero eigenfrequencies $\omega^2_0=0$ of the fundamental radial oscillation mode occur.  The sign of the separation in central density depends on the charge model; this separation  also has a dependence on $\lambda$ such that  increasing $\lambda$ (which can occur for either large color superconductivity or small strange quark mass) tends to decrease this separation size for the first and third classes of charge models monotonically. 
Moreover, for the second and third classes of charge models, we manage to numerically and analytically identify a new kind of stellar structure with a zero central pressure but with a finite radius and mass.   All the calculations and analysis are performed in a general dimensionless rescaling approach so that the results are independent of explicit values of dimensional parameters.

\end{abstract}
\maketitle
\clearpage
\section{Introduction}

One of the key challenges of strong interaction physics is that of understanding matter at very high densities \cite{Pasechnik:2016wkt}.  
This regime can be explored at high temperatures by the future NICA and FAIR experiments, whereas at low temperatures
the behavior of high density matter determines the most compact stable stars. Empirically it is known that matter can bind at nuclear densities to form neutron stars.  However, it is theoretically possible to exceed this limit, in which case the most compact stable stars would be  described in terms of deconfined quark degrees of freedom.

 The Bodmer-Witten-Terazawa hypothesis~\cite{Bodmer:1971we, Witten, Terazawa:1979hq} suggests that quark matter with comparable amounts of $u, \,d, \,s$ quarks, also called strange quark matter (SQM), might be the ground state of baryonic matter at the low (zero) temperature and pressure.  This view has recently been challenged by a  study~\cite{Holdom:2017gdc}  demonstrating  that $u, d$ quark matter ($ud$QM) is, in general, more stable than SQM, and indeed can be more stable than the ordinary nuclear matter at a sufficiently large baryon number beyond the periodic table. Such bulk absolute stability allows the possibility of quark stars consisting of $ud$QM, or up-down quark stars, in addition to strange quark stars consisting of SQM.


Interacting quark matter includes  effects from strong interactions that can modify its behavior, such as one-gluon exchange or color superconductivity. One-gluon exchange induces perturbative quantum chromodynamics (pQCD) corrections to the free energy~\cite{Farhi:1984qu,Fraga:2001id,Fraga:2013qra}. Furthermore, a spin-0 Cooper-pair condensate antisymmetric in color-flavor space is expected  to lower the energy~\cite{Alford:1998mk,Rajagopal:2000ff,Lugones:2002va}, resulting in two-flavor color superconductivity, where $u$ quarks pair with $d$ quarks. This is  conventionally termed ``2SC" /``2SC+s" without/with  strange quarks, or the color-flavor locking (CFL) phase in which $u,d,s$ quarks pair with each other antisymmetrically. 

Taking into account color superconductivity and pQCD corrections generally yields predictions of  larger masses and radii for quark stars, due to their stiff 
equation of state (EOS) compared to those of neutron stars and ``normal" quark stars in which such interactions are neglected~\cite{Zhang:2019mqb} . Such interacting quark stars have been shown to have regions of  parameter space that match current astrophysical constraints~\cite{Ren:2020tll,Zhang:2020jmb} and are able to produce gravitational-wave echoes~\cite{Zhang:2021fla}.  


Compact stars are conventionally assumed to be (net) charge-neutral (or nearly neutral) objects. For cold stars, a crude estimation of the maximum charge can be obtained from a Newtonian approximation, which compares the gravity of an individual particle $GMm/R^2$ to its electrostatic force $GQq/R^2$ at the star surface, yielding a net charge of approximately 100 Coulombs per solar mass. However, for compact stars where the degeneracy pressure of the particles also balances the gravitational pull, 
relativistic effects and high densities must be taken into account ~\cite{Bekenstein:1971ej}. Compact stars with a certain charge density profile can hold a huge net charge (of order $10^{20}$C) determined from the global force balance conditions~\cite{Ray:2003gt}. This large charge is well beyond the stability bound ($100$ C) set by the Newtonian approximation for an individual charged particle at the star surface. In this sense, as Ref.~\cite{Ray:2003gt} argued, such a star would be short lived, since individual charged particles would rapidly be ejected from it, upon which it would collapse to a black hole. 

This argument likely does not  apply to quark stars since they are not composed of neutral objects but rather a gas of quarks and gluons.  Considering the abundance of positively charged (up) quarks and the associated strong interaction effects, the possibility that charged quark stars exist remains an interesting open question.
Indeed, there have been a number of studies on charged quark stars~\cite{Negreiros:2009fd,Arbanil:2015uoa, Panotopoulos:2019wsy,Goncalves:2020joq,Panotopoulos:2021cxu,Jasim:2021cft}. However, none of them unified   pQCD  corrections and color superconductivity effects. Furthermore, their analysis and results heavily depended on some \textit{ad hoc} values of several empirical parameters, like the bag constant, strange quark mass, superconducting gap, flavor composition, etc. 

The oscillation modes of a compact star encode rich information about its internal structure and thermodynamic properties.  Such studies are usually termed neutron-star seismology or astroseismology~\cite{Sagun:2020qvc,Andersson:2021qdq}. Of particular interest to our study is the fundamental mode of radial oscillations, since a positive square of such an eigenfrequency  indicates the star is stable against radial perturbations.

Previously we established a general parametrization of the quark matter  EOS  that included  pQCD corrections and   color superconductivity~\cite{Zhang:2020jmb}. Somewhat remarkably, we found that the EOS could be rescaled into a dimensionless form  depending only on a single parameter characterizing the relative size of strong interaction effects. In this paper we employ this approach to  explore the stellar structure and radial stability of charged quark stars with   different charge configurations. 

The unified EOS of interacting quark matter relating the radial pressure $P$ to the mass density $\rho$ is~\cite{Zhang:2020jmb}:
\be
P=\frac{1}{3}(\rho-4B_{\rm eff})+ \frac{4\lambda^2}{9\pi^2}\left(-1+\rm sgn(\lambda)\sqrt{1+3\pi^2 \frac{(\rho-B_{\rm eff})}{\lambda^2}}\right)
\label{eos_tot}
\ee
where $B_{\rm eff}$ is the effective bag constant that accounts for the nonperturbative contribution from the QCD vacuum and
\be
\lambda=\frac{\xi_{2a} \Delta^2-\xi_{2b} m_s^2}{\sqrt{\xi_4 a_4}}.
\label{lam}
\ee
The gap parameter $\Delta$ represents the color-superconducting gap, and $m_s$ denotes the strange quark mass. $a_4$ represents the pQCD corrections from one-gluon exchange for gluon interaction to $O(\alpha_s^2)$ order, which can vary from $a_4=1$, corresponding to a vanishing pQCD correction, to very small values where these corrections become large.   
Note that $\rm sgn(\lambda)$ represents the sign of $\lambda$, which is positive as long as $\Delta^2/m_s^2>\xi_{2b}/\xi_{2a}$. 
The constant coefficients in $\lambda$ are
\begin{align}
(\xi_4,\xi_{2a}, \xi_{2b}) = \left\{ \begin{array} {ll}
(( \left(\frac{1}{3}\right)^{\frac{4}{3}}+ \left(\frac{2}{3}\right)^{\frac{4}{3}})^{-3},1,0) & \textrm{2SC phase}\\
(3,1,3/4) & \textrm{2SC+s phase}\\
(3,3,3/4)&   \textrm{CFL phase}
\end{array}
\nonumber
\right.
\end{align}
characterizing the possible phases of color superconductivity. 
Furthermore, the dimensionless rescaling
\be
\bar{\rho}=\frac{\rho}{4\,B_{\rm eff}}, \,\, \bar{P}=\frac{P}{4\,B_{\rm eff}},  \,\,
\label{rescaling_prho}
\ee
and 
\be
 \bar{\lambda}=\frac{\lambda^2}{4B_{\rm eff}}= \frac{(\xi_{2a} \Delta^2-\xi_{2b} m_s^2)^2}{4\,B_{\rm eff}\xi_4 a_4},
 \label{rescaling_lam}
\ee
further removes the $B_{\rm eff}$ parameter, 
so that the EOS Eq.~(\ref{eos_tot}) reduces to the dimensionless form 
\be
\bar{P}=\frac{1}{3}(\bar{\rho}-1)+ \frac{4}{9\pi^2}\bar{\lambda} \left(-1+\rm sgn(\lambda)\sqrt{1+\frac{3\pi^2}{\bar{\lambda}} {(\bar{\rho}-\frac{1}{4})}}\right),
\label{eos_p}
\ee
or conversely
\be
\bar{\rho}=3\bar{P}+1- \frac{4}{\pi^2}\bar{\lambda} \left(-1+\text{sgn}(\lambda)\sqrt{1+\frac{\pi^2}{\bar{\lambda}} {(\bar{P}+\frac{1}{4})}}\right).
\label{eos_rho}
\ee
One can easily see  that a stiffer EOS   results from   increasingly positive values of $\lambda$, which corresponds to a larger (smaller) $\bar{\lambda}$ when the sign of $\lambda$ is positive (negative).
Referring to Eq. (\ref{rescaling_lam}),  a larger pQCD (i.e., smaller $a_4$) always maps to a larger $\bar{\lambda}$. However, a larger color superconducting effect (i.e., larger $\Delta$) or a smaller $m_s$ only maps to a larger $\bar{\lambda}$ when $\lambda$ is positive. 

 For positive $\lambda$,  taking the limit where $\bar{\lambda}$ goes extremely large, Eq.~(\ref{eos_p}) approaches the special form
\be
\bar{P}\vert_{\bar{\lambda}\to \infty}=\bar{\rho}-\frac{1}{2}, 
\label{eos_infty}
\ee
which is equivalent to $p={\rho}-2B_{\rm eff}$ after scaling back with Eq.~(\ref{rescaling_prho}). However, for negative $\lambda$, taking the EOS to $\bar{\lambda}\to \infty$ limit does not yield a finite form. 
As $\bar{\lambda}\to0$, Eq.~(\ref{eos_p}) reduces to the conventional noninteracting rescaled quark matter EOS  $\bar{P}=(\bar{\rho}-1)/3$.
\section{Rescaled charged TOV equations}

The stellar structure of charged interacting quark stars is obtained by solving the Einstein-Maxwell equations for
the metric
\be \label{metric}
ds^2 = -  e^{2\nu} dt^2 + e^{2\Lambda} dr^2 + r^2 \left(d\theta^2 + \sin^2\theta d\phi^2\right)
\ee
where $\nu$ and $\Lambda$ are functions of $(r,t)$.  The equilibrium solution is obtained from the time-independent case,
yielding the 
Tolman-Oppenheimer-Volkov (TOV) equations~\cite{Oppenheimer:1939ne,Tolman:1939jz}  with   charge effects included~\cite{Ray:2003gt}
\begin{align}
  \frac{\mathrm{d}q}{\mathrm{d}r} & {} = 4\pi r^2 \rho_e e^{\Lambda} \,, \label{TOV1} \\
  \frac{\mathrm{d}m}{\mathrm{d}r} & {} = 4\pi r^2 \rho  + \frac{q}{r}\frac{\mathrm{d}q}{\mathrm{d}r} \,,  \label{TOV2} \\
  \frac{\mathrm{d}P}{\mathrm{d}r} & {} = -(\rho + P)\left(4\pi r P + \frac{m}{r^2} - \frac{q^2}{r^3}\right)e^{2\Lambda}  
  + \frac{q}{4\pi r^4}\frac{\mathrm{d}q}{\mathrm{d}r}   \label{TOV3} \\
   \frac{\mathrm{d}\nu}{\mathrm{d}r}  & {} = -\frac{1}{\rho + P}\left(\frac{\mathrm{d}P}{\mathrm{d}r} - \frac{q}{4\pi r^4}\frac{\mathrm{d}q}{\mathrm{d}r}\right) \,,  \label{TOV4}
  \end{align}
where $q(r)$ and $m(r)$,  respectively, represent the charge and mass within radius $r$.  $\rho_e(r)$ is the electric charge density at $r$, and
\begin{equation}
  \label{chargedTOV}
  e^{-2\Lambda(r)} = 1 - \frac{2m(r)}{r} + \frac{q(r)^2}{r^2}\,.
\end{equation} 

To solve the charged TOV equations  \eqref{TOV1} --  \eqref{TOV4}  at a given
 central mass density $\rho_c$ we employ the boundary conditions
\begin{equation}
  \label{eq:r0}
  q(0) = m(0) = 0\,,\qquad \rho(0) = \rho_{\mathrm{c}}\,,\qquad  \nu(R) = -\Lambda(R)\,,
  \end{equation}
where the star's radius $R$ is determined by the condition $P(R)=0$. The relations $m(R)=M$ and $q(R)=Q$ determine the mass $M$ and  total charge $Q$ of the star. 
We observe that we can do the following rescaling
\be
 \bar{m}=m{\sqrt{4\,B_{\rm eff}}}, \quad \bar{r}={r}{\sqrt{4\,B_{\rm eff}}}, \,\,
\label{rescaling_mr}
\ee
and
\be
\bar{q}={q}{\sqrt{4\,B_{\rm eff}}}, \quad \bar{\rho}_e=\frac{\rho_e}{4\,B_{\rm eff}},
\label{rescaling_charge}
\ee
to cast the charged TOV equations into a dimensionless form, in which barred symbols in  \eqref{TOV1} --  \eqref{TOV4}   replace unbarred ones.  

We shall investigate three benchmark models that are most widely used in the literature~\cite{Ray:2003gt,Negreiros:2009fd,Arbanil:2015uoa, Panotopoulos:2019wsy,Goncalves:2020joq,Panotopoulos:2021cxu}. 
 \begin{itemize}
\item  Model A: {\it charge density is proportional to  energy density}
\begin{equation}
  \label{eq:qalpha}
  \rho_e = \alpha\rho\,,
\end{equation}
where $\alpha$ is a dimensionless parameter in the range $0\leq \alpha \leq 1$. After  rescaling using (\ref{rescaling_prho}) and  (\ref{rescaling_charge}), we obtain the dimensionless relation
\begin{equation}
  \label{eq:qalpha}
  \bar{\rho}_e = \alpha \bar{\rho}\,
  \end{equation}
\item Model B: {\it charge is proportional to  spatial volume}
\begin{equation}
  \label{eq:qbeta}
  q(r) = Q\left(\frac{r}{R}\right)^3 \equiv \beta r^3\,,
  \end{equation}
  which reads 
  \be
\bar{q}(\bar{r})=\bar{\beta} \bar{r}^3.
\ee
in dimensionless form, where 
  \be
  \bar{\beta}=\frac{\beta}{4B_{\rm eff}}.
  \label{rescaling_beta}
\ee

\item Model C: {\it fixed total charge}\\ 
Fixing the total charge to be $Q$, this model corresponds fixing the dimensionless quantity
\be
\bar{Q}={Q}{\sqrt{4\,B_{\rm eff}}}
\ee
using~(\ref{rescaling_charge}). 

\end{itemize}

\section{Radial stability}

To investigate radial stability, we assume that a fluid element is displaced from its equilibrium position $r$ to $r+\delta r$, and that
such a perturbation has harmonic time dependence $e^{i\omega t}$.  
The equation for infinitesimal radial oscillations of a spherical object is given by~\cite{Brillante:2014lwa}

\begin{equation}
  \label{eq:eta}
  \frac{\mathrm{d}u}{\mathrm{d}r} = \frac{ \eta}{\cal{P}} \,, 
\end{equation}
and
\begin{equation}
  \label{eq:detadr}
  \frac{\mathrm{d} \eta}{\mathrm{d} r} = -[{\cal{Q}} + \omega^2{\cal{W}}]u \,.
\end{equation}
where $u = r^2 e^{-\nu(r)} \delta r$ is the renormalized displacement function~\cite{Brillante:2014lwa}  and
\begin{align}
  \label{eq:pqw}
  \cal{P} & {} = e^{\Lambda+3\nu}r^{-2}\gamma P\,, \\
  \cal{Q} & {} = (\rho + P)r^{-2}e^{\Lambda + 3\nu}\left[\nu'\left(\nu' - 4r^{-1}\right) - (8\pi P  + r^{-4}q^2)e^{2\Lambda}\right] \,,\\
  \cal{W} & {} = e^{3\Lambda+\nu}r^{-2}(\rho + P)\,,
\end{align}
with
\be
 \gamma=\frac{\rho+P}{P}\frac{dP}{d\rho}
 \ee 
as the adiabatic index.
The initial conditions are $\eta(0) = 1$ and correspondingly $u(0) = r^3/(3{\cal{P}}(0))$. 
A series of eigenfrequencies $\omega_n^2$ can be obtained by solving \eqref{eq:eta} and \eqref{eq:detadr} 
via the shooting method, satisfying the  boundary condition 
\begin{equation}
  \label{eq:dudr}  
  \frac{\mathrm{d}u}{\mathrm{d}r}\bigg\vert_{r=R} = \eta(R) = 0\,.
\end{equation}
Radial stability can be determined by requiring a non-negative fundamental mode eigenfrequency $\omega_0^2\geq0$. For uncharged compact stars, the point where $\omega_0^2\geq0$ coincides with the condition $\partial M/ \partial \rho_C\geq0$, implying the zero fundamental eigenfrequency point coincides with the maximum mass point~\cite{Glendenning}. However, this coincidence does not hold for the charged case~\cite{Arbanil:2015uoa, Goncalves:2020joq}.  In the following calculations, we rescale all the variables in this radial stability analysis into their dimensionless forms introduced previously (simply replacing the unbarred symbols with barred ones), with the rescaled dimensionless eigenfrequencies being
$\bar{\omega}_n=\omega_n/\sqrt{4 B_{\rm eff}}.$


\section{results for Positive $\lambda$}
Here we present the results of the stellar structure and the radial stability for charged interacting quark stars with a positive $\lambda$.
\subsection{Model A}

\begin{figure}[h]
 \centering
 \includegraphics[width=8.15 cm]{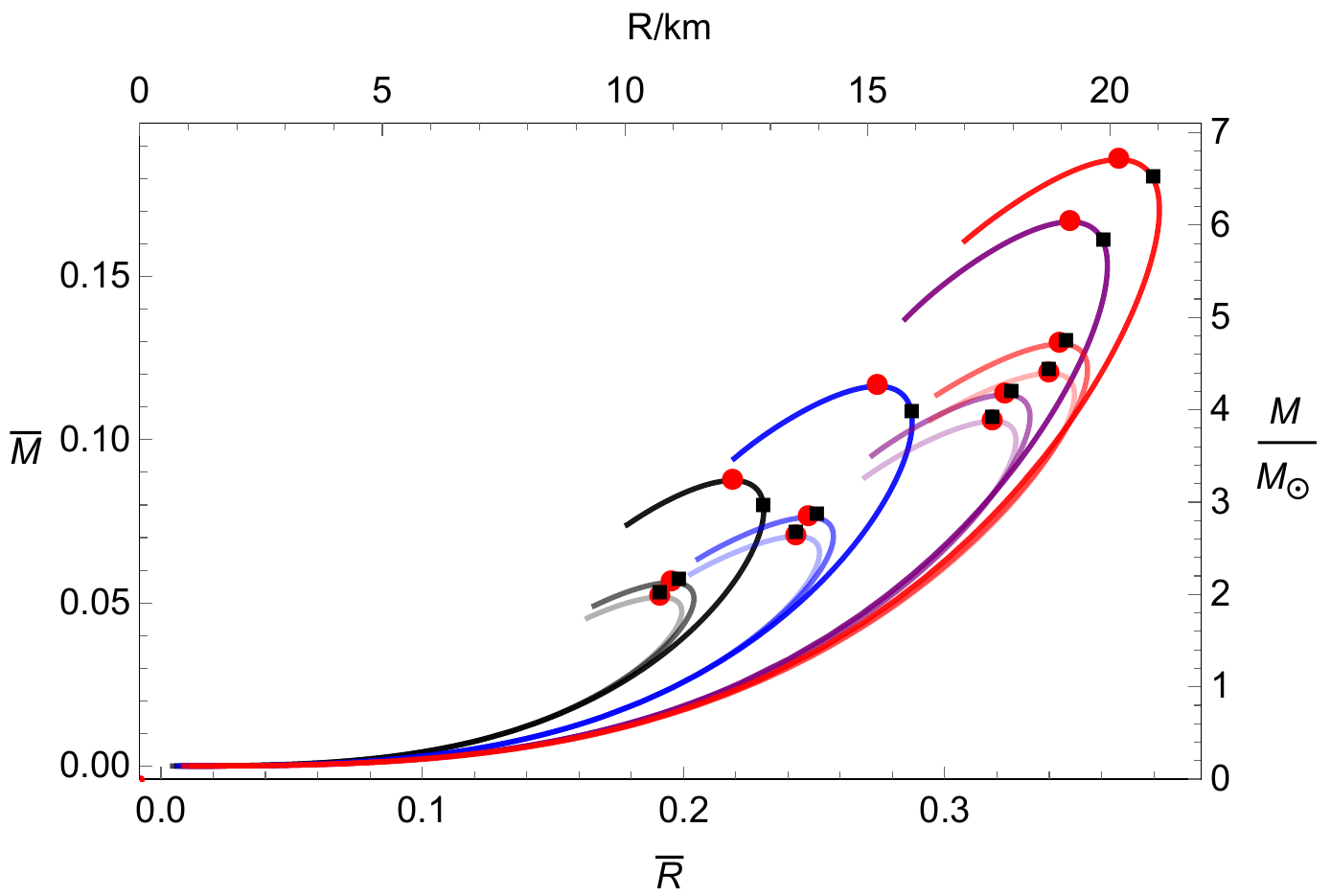}  
\includegraphics[width=8.15cm]{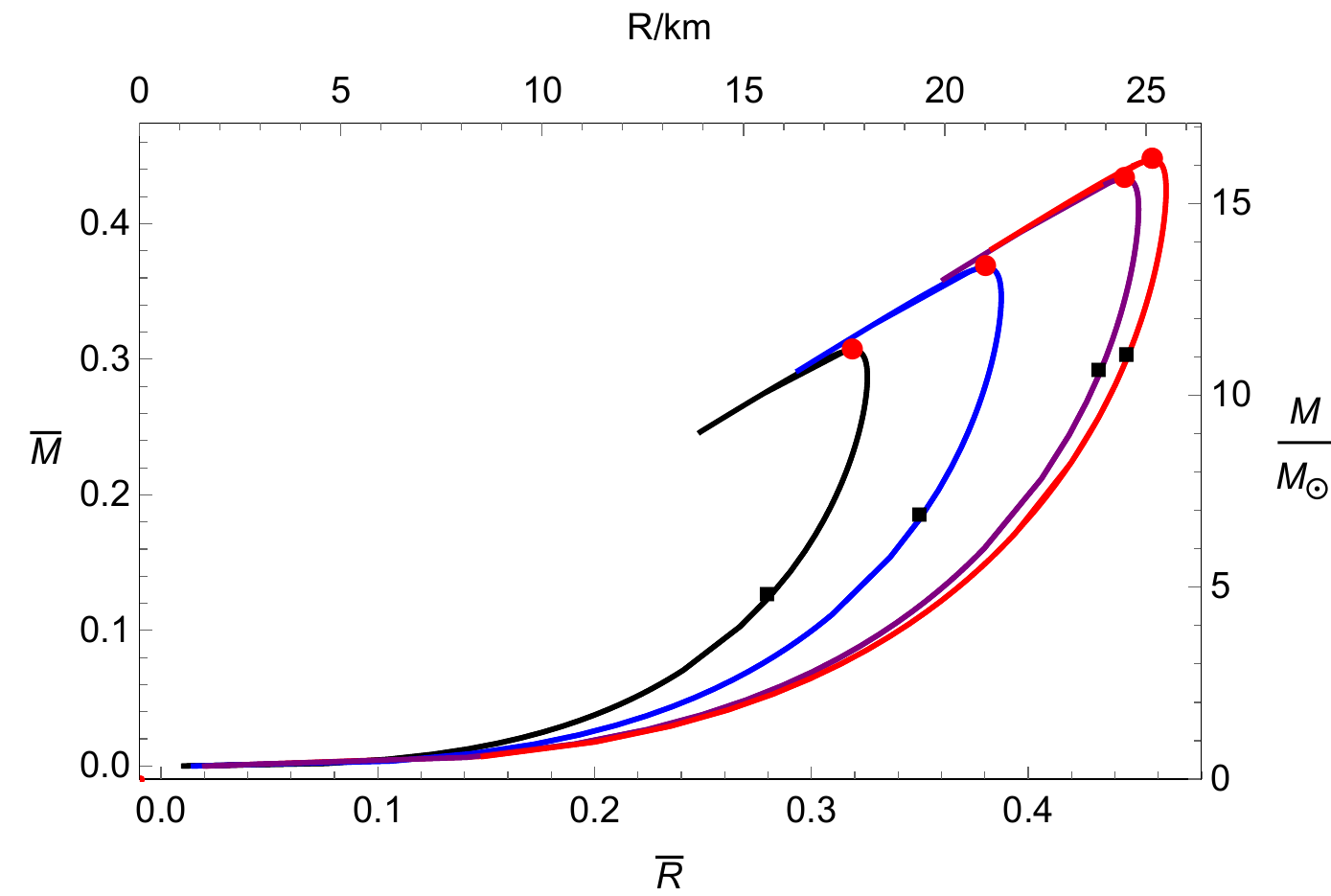}  
   \includegraphics[width=8.15cm]{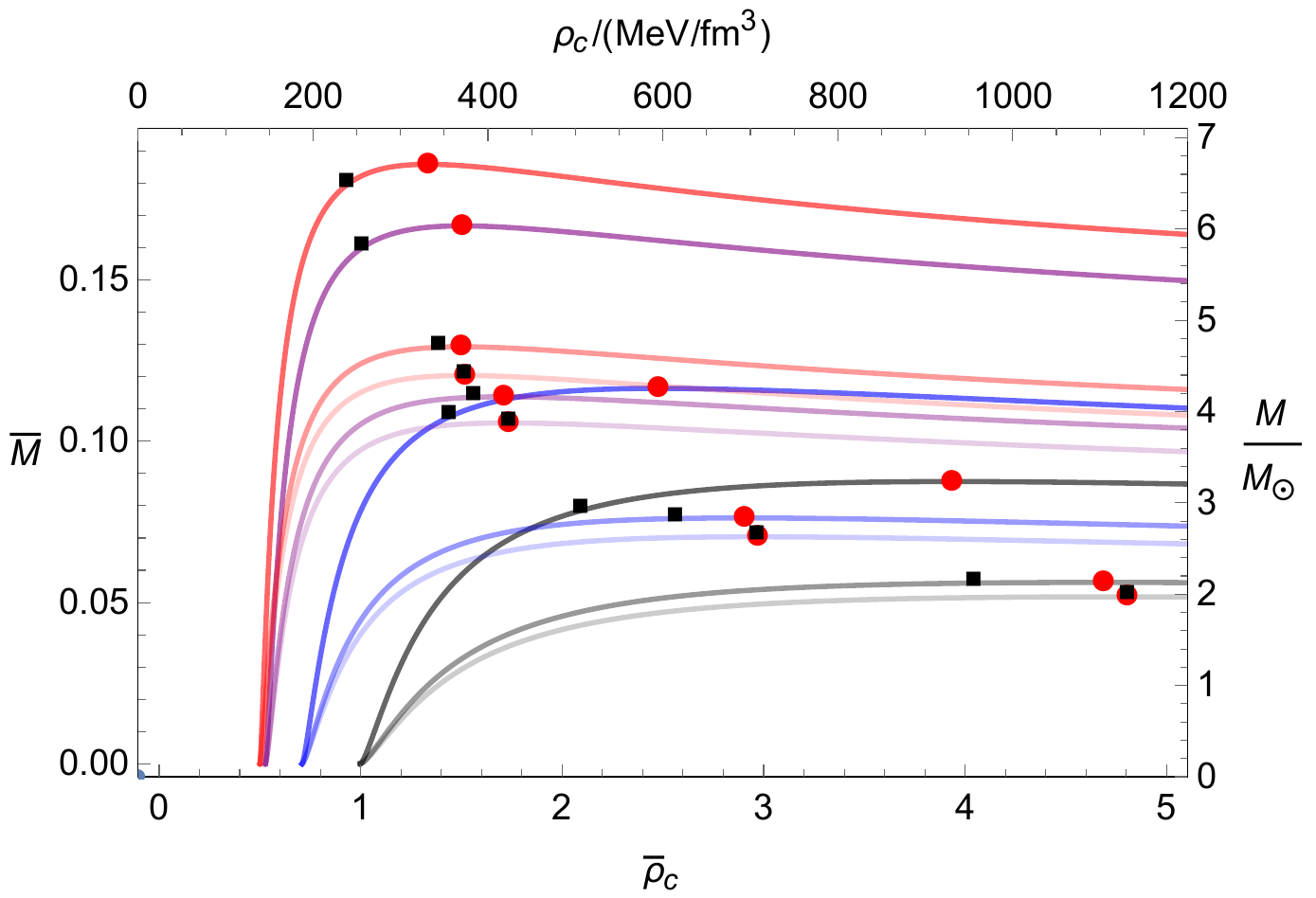}  
  \includegraphics[width=8.15cm]{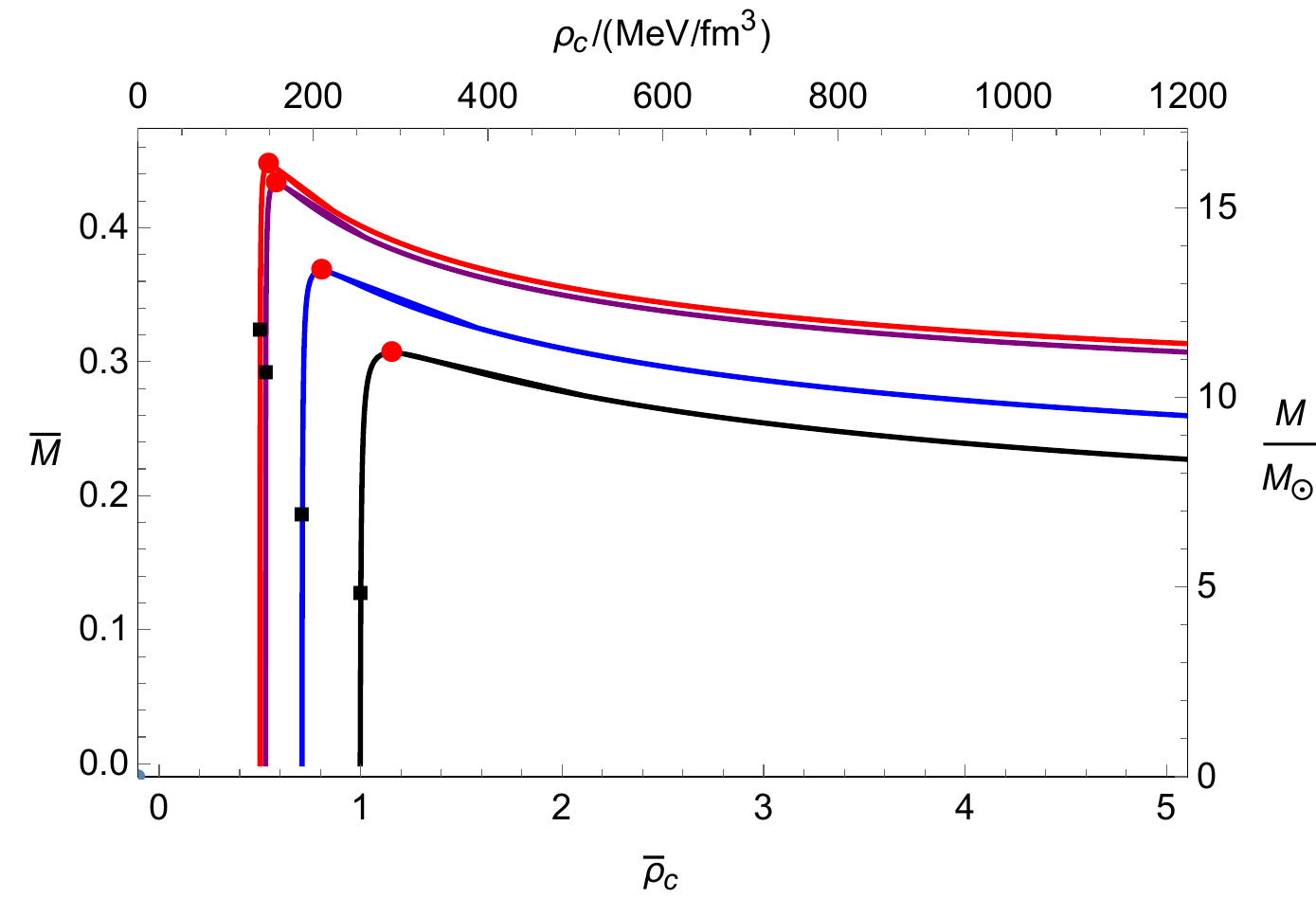}  
\caption{Plots of $\bar{M}$ vs. $\bar{R}$ (upper graphs) and $\bar{\rho}_c$ (lower graphs) of charged interacting quark stars in model A for (left) $\alpha=(0, 0.3, 0.7)$ and (right) $\alpha=0.999$. The right and top axes in each plot are the corresponding dimensional parameters with $B_{\rm eff}=60\, \rm MeV/fm^3$ for illustration.
The  black, blue, purple, and red curves respectively denote   $\bar{\lambda}=(0, 0.5 , 10, \infty)$ with the sign of $\lambda$ being positive. Darker shades correspond to increasing values of $\alpha$.
The solid dots denote the maximum mass configurations, with the filled squares denoting where $\bar{\omega}^2_0=0$.}
   \label{rescaledMR_caseA}
\end{figure}

Plots of mass in terms of radius and $\bar{\rho}_c$ of interacting quark stars with model-A-type charge configurations are shown in Fig.~\ref{rescaledMR_caseA}, where the red dots and black squares, respectively, indicate the points of maximum mass  and   zero eigenfrequency of the fundamental radial oscillation mode  for each case.  We plot results for four different values of $\alpha$.

We see that a larger $\alpha$ and $\bar{\lambda}$ result in a larger mass and radius in the $\bar{M}$-$\bar{R}$ plots, and lead to a stiffer slope for the $\bar{M}$-$\bar{\rho}_c$ curves yielding a smaller central density at the maximum mass point. 
Note that the location of $\bar{\rho}_c$ at $\bar{M}=0$ is predicted by Eq.~(\ref{eos_rho}) with the pressure $P$ set to zero.
We also see that if $\alpha \neq 0$ the maximum mass points and the  $\bar{\omega}^2_0=0$ points are not coincident,  with the latter occurring for smaller central densities. This is particularly evident in the lower two diagrams of
Fig.~\ref{rescaledMR_caseA}, where we see that the stability point $\bar{\omega}^2_0=0$ occurs for smaller values of 
$\bar{\rho}_c$ than the maximum mass point, with the size of the separation in mass between these two points becoming larger with increasing $\alpha$, and the separation size in density becoming smaller with increasing $\bar{\lambda}$.

\subsection{Model B}

\begin{figure}[h]
 \centering
 \includegraphics[width=8.15cm]{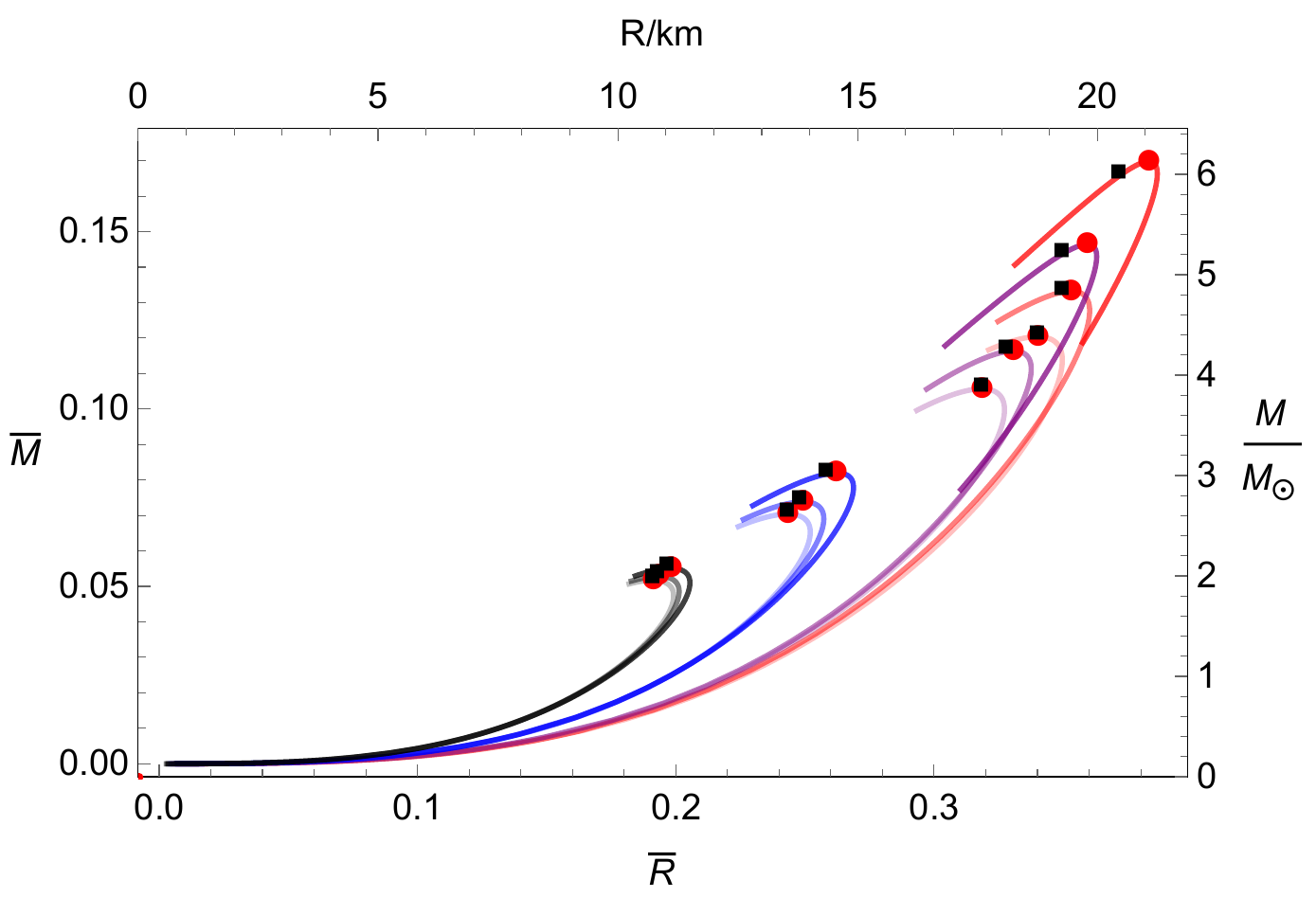}  
  \includegraphics[width=8.15cm]{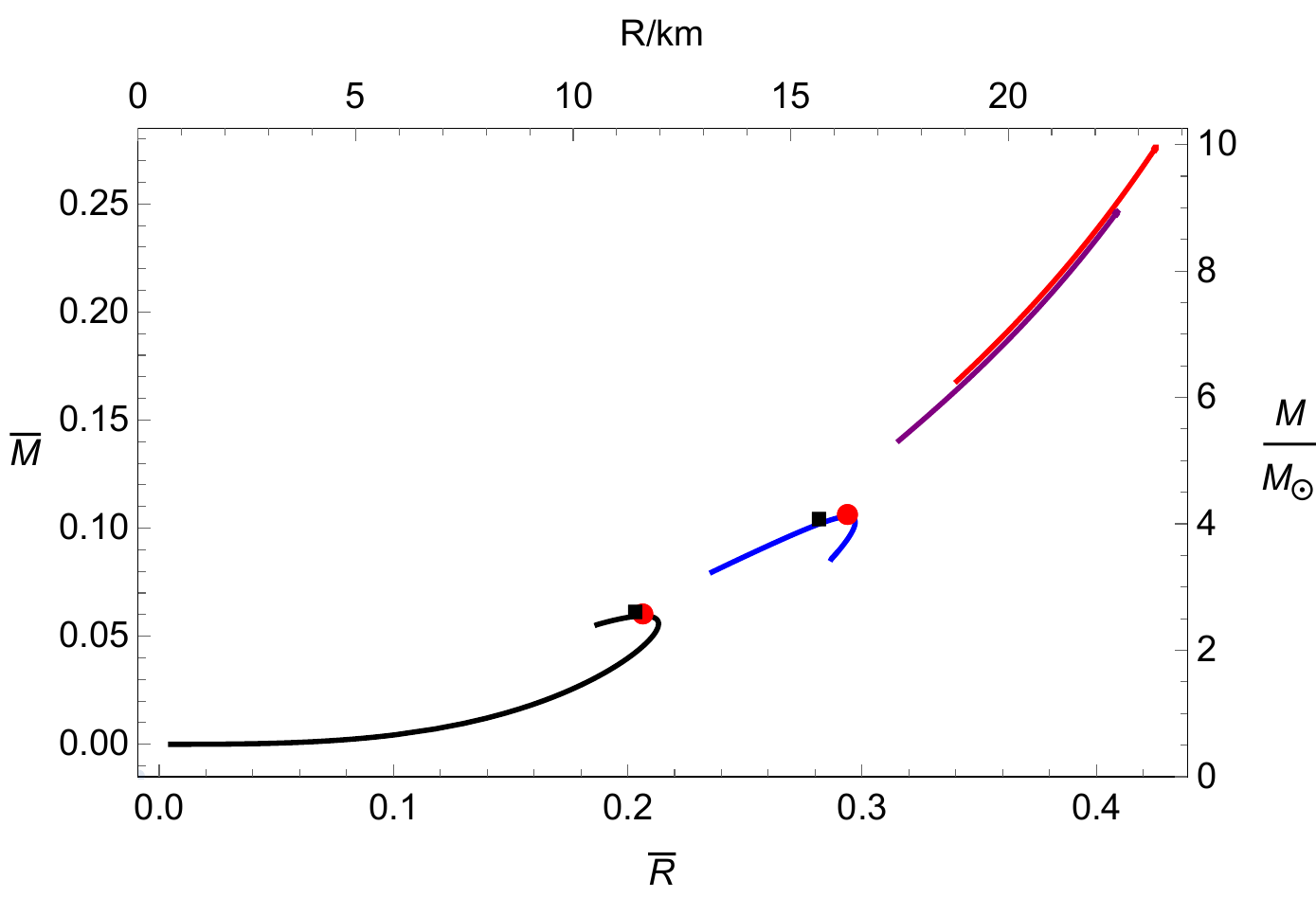}  
   \includegraphics[width=8.15cm]{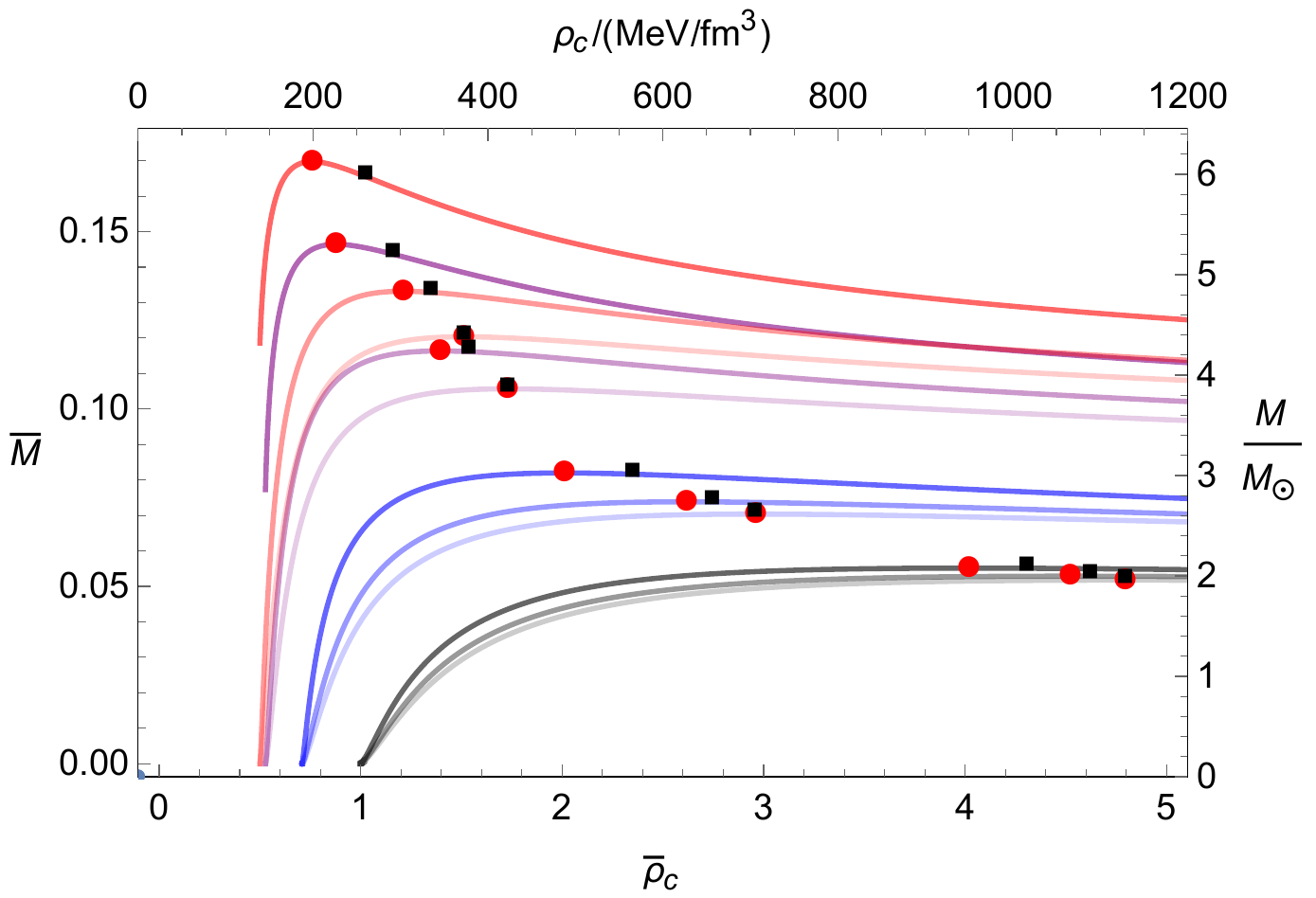}  
  \includegraphics[width=8.15cm]{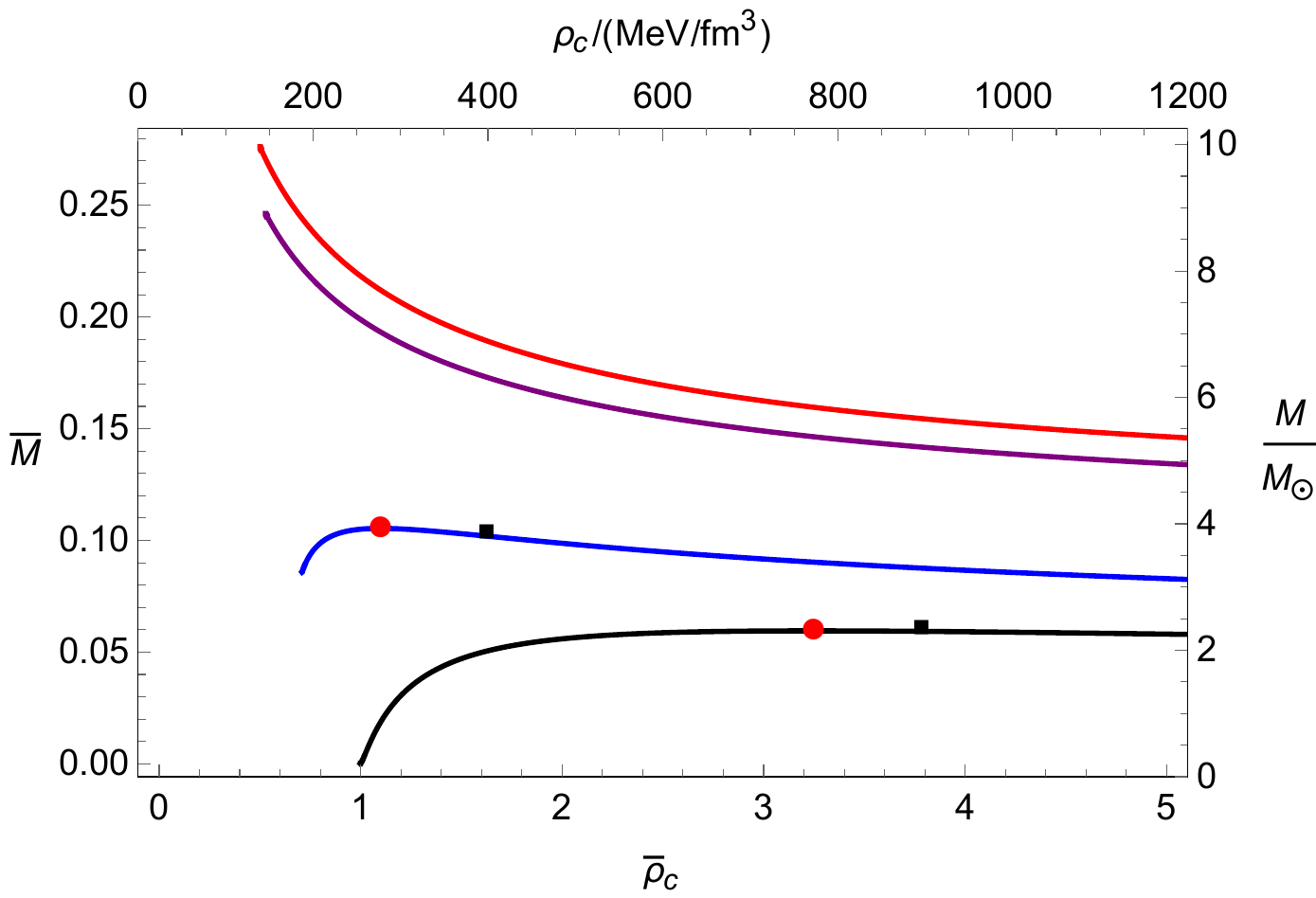}  
\caption{Plots of $\bar{M}$ vs. $\bar{R}$ (upper graphs) and $\bar{\rho}_c$ (lower graphs) of charged interacting quark stars in model B for (left) $\bar{\beta}=(0,1.5, 2.5)$   and (right) $\bar{\beta}=3.5$. The right and top axes in each plot are the corresponding dimensional parameters with $B_{\rm eff}=60\, \rm MeV/fm^3$ for illustration.  The  black, blue, purple, and  red curves respectively denote  $\bar{\lambda}=(0, 0.5 , 10, \infty)$ with the sign of $\lambda$ being positive. Darker shades correspond to increasing values of $\bar{\beta}$.
The solid dots denote the maximum mass configurations, with the filled squares denoting where $\bar{\omega}^2_0=0$. In the right diagrams, the configurations denoted by the purple and red curves are radially unstable for all central densities.}
   \label{rescaledMR_caseB}
\end{figure}

We depict the mass/radii curves for   interacting quark stars with model-B-type charge  configurations   in Fig.~\ref{rescaledMR_caseB}. As before, both increasing $\bar{\lambda}$ and increasing charge enlarge the mass and radius of the star, and lead to a stiffer slope for the $\bar{M}$-$\bar{\rho}_c$ curves  resulting in a smaller central density at the maximum mass point.   However we see here that  the stability point $\bar{\omega}^2_0=0$ now occurs for  larger values of 
$\bar{\rho}_c$ than the maximum mass point, with the sizes of  the mass separation and the density separation between these two points both becoming larger with increasing $\bar{\beta}$.  Interestingly, we find the density separation size first gets  rapidly larger, then slowly becomes   smaller   as $\bar{\lambda}$ increases. For sufficiently large $\bar{\beta}$ and $\bar{\lambda}$ there are no stability points, and the configurations are radially unstable for all central densities, as shown in the upper two curves in the right diagrams of Fig.~\ref{rescaledMR_caseB}.

\begin{figure}[h]
 \centering
 \includegraphics[width=8.33cm]{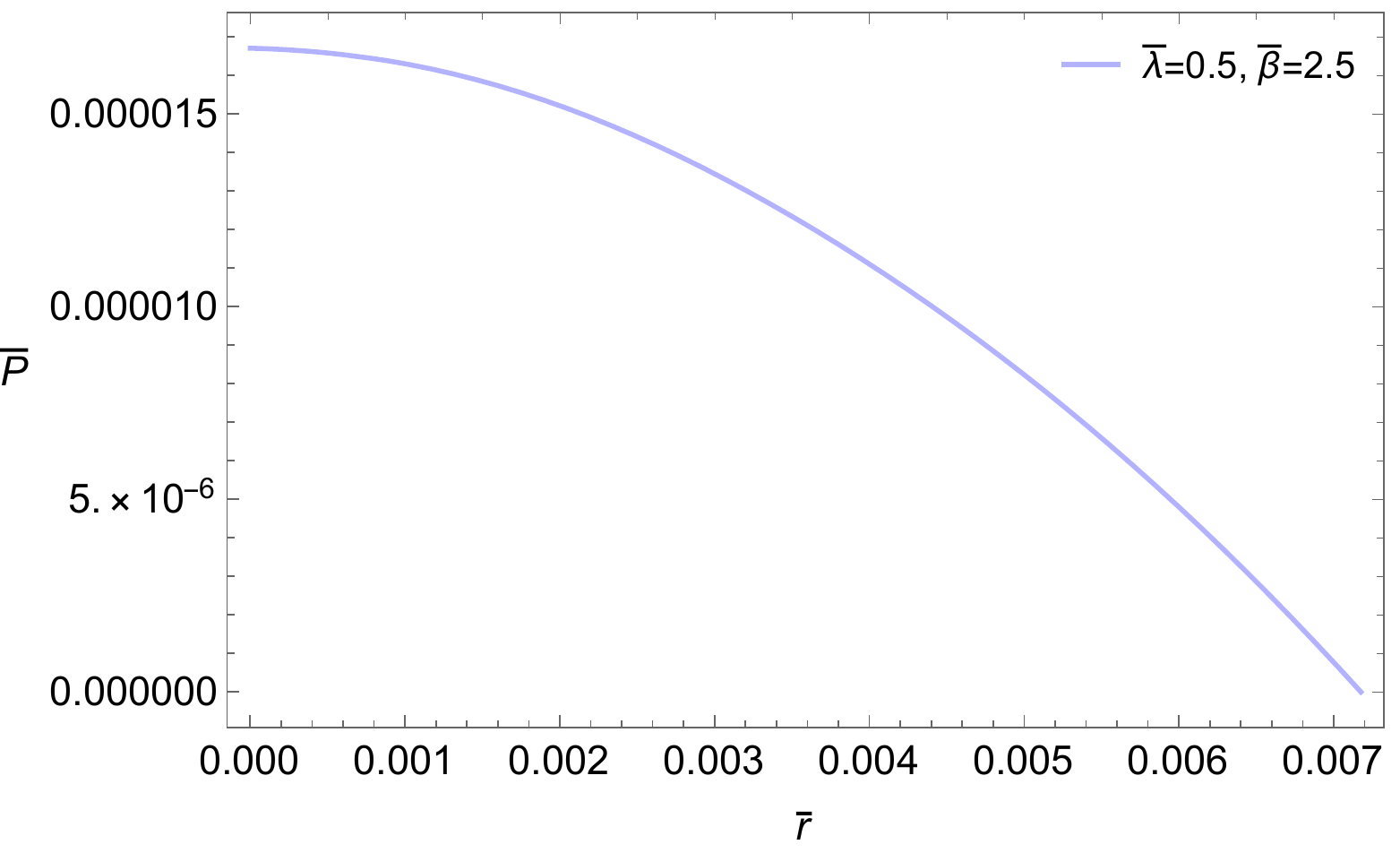}  
  \includegraphics[width=8cm]{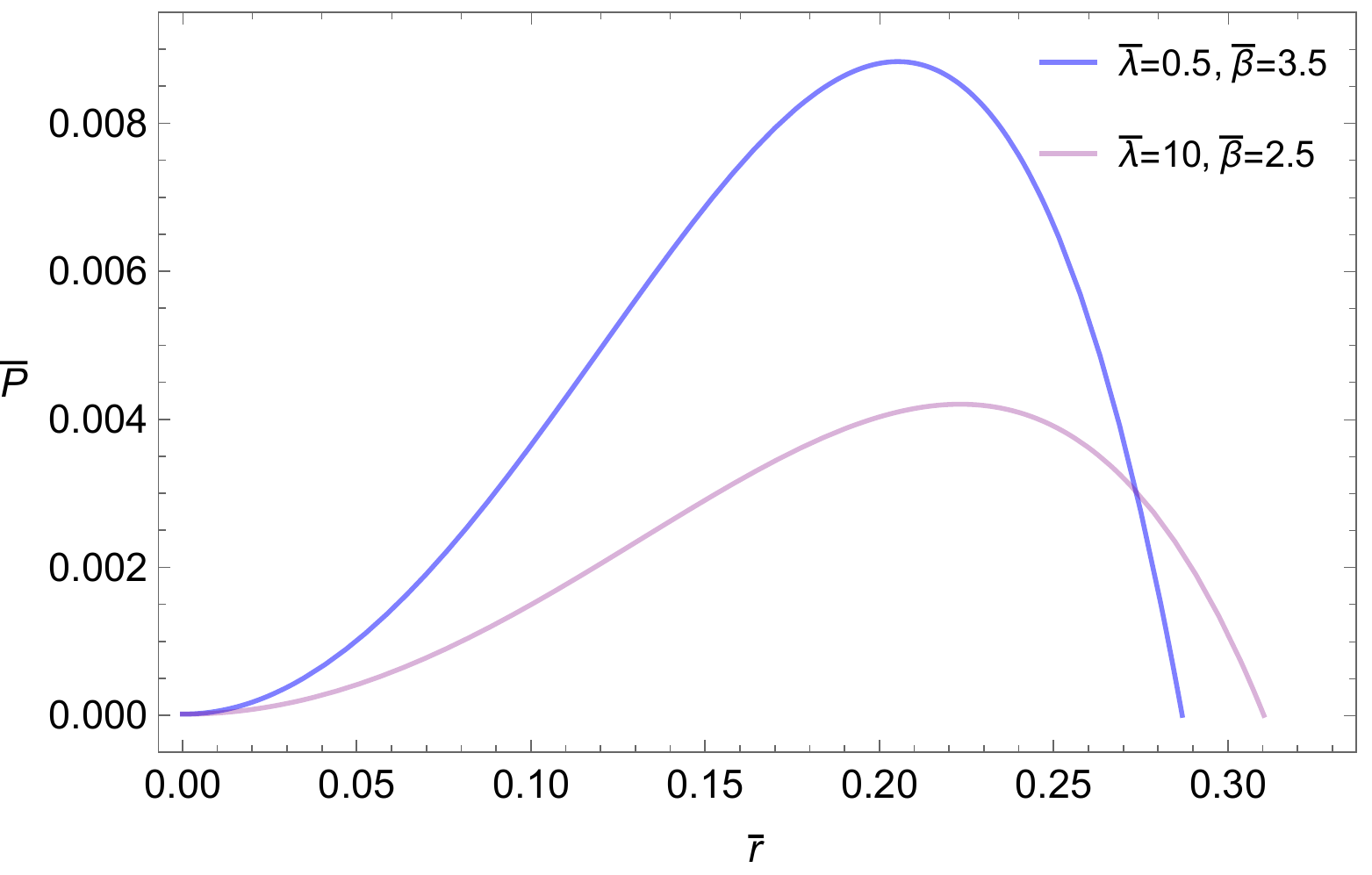}  
\caption{$\bar{P}$-$\bar{r}$ plots of charged interacting quark stars in model B for (a) $(\bar{\lambda}, \bar{\beta})=(0.5,2.5)$ and (b) $(\bar{\lambda}, \bar{\beta})=(0.5, 3.5), (10, 2.5)$ with a tiny rescaled central pressure $\bar{P}=0.00001671$.}
   \label{rescaledPr_caseB}
\end{figure}

Another interesting feature for large $\bar{\beta}$ and large $\bar{\lambda}$ configurations  is that the stellar structure with nearly zero central pressure has non-zero radii and masses (i.e., $\bar{M}$-$\bar{R}$ curves do not start from the origin). This can be seen from the pressure profile plots shown in Fig.~\ref{rescaledPr_caseB}. In contrast to normal cases where $\bar{P}(\bar{r})$ decreases over increasing $\bar{r}$ monotonically as shown in Fig.~\ref{rescaledPr_caseB}a, we see in Fig.~\ref{rescaledPr_caseB}b that for some large $(\bar{\lambda}, \bar{\beta})$ sets, $\bar{P}(\bar{r})$ grows first before it decreases over $r$, resulting in a finite radius even at a tiny central pressure. This is because the mass density or the corresponding pressure profile has to grow (over $\bar{r}$) first to counterbalance the effect of a rapidly growing charge density when $\bar{\beta}$ is large enough.
Therefore, the critical value of $\beta$ for which such a transition at near-zero central pressure takes place is a result of the competition between  gravitational and electrostatic forces at the origin. This can be described as
\be
M_{r\to0}=Q_{r\to0} \Rightarrow \rho_0\frac{4\pi}{3}r^3 =\beta_c r^3\Rightarrow \beta_c=\frac{4\pi}{3}\rho_0
\ee
where $\rho_0=\rho(r=0)$. Rescaling this into dimensionless form using~(\ref{rescaling_prho}) and  (\ref{rescaling_beta})
and inserting it into the rescaled EOS Eq.~(\ref{eos_rho}), we obtain 
\be
\bar{\beta}_c={ }\frac{4\pi}{3} \left(3\bar{P}_0+1- \frac{4}{\pi^2}\bar{\lambda} \left(-1+\sqrt{1+\frac{\pi^2}{\bar{\lambda}} {(\bar{P}_0+\frac{1}{4})}}\right) \right).
\label{betac}
\ee
 As the central pressure $\bar{P}_0\to 0$ we have 
\be
\bar{\beta}_{c0}= \frac{4\pi}{3}\left(1- \frac{1}{1+\sqrt{1+\pi^2/(4\bar{\lambda})}}\right)
\label{betac0}
\ee
and we see that  larger $\bar{\lambda}$  results in a smaller $\bar{\beta}_{c0}$, with the smallest  value $\bar{\beta}_{c0}= 2\pi/3\approx 2.094$ achieved for $\bar{\lambda}\to \infty$, and the largest value $\bar{\beta}_{c0}=4\pi/3\approx 4.189$   achieved for $\bar{\lambda}=0$.
 This explains the finite $(\bar{M}, \bar{R})$ values at zero central pressure for large $\bar{\beta}$ and large $\bar{\lambda}$ shown in Fig.~\ref{rescaledMR_caseB} and Fig.~\ref{rescaledPr_caseB}.

For very large $\bar{\beta}$ and $\bar{\lambda}$ values,  we find configurations radially unstable for all central densities, as denoted by the purple and red curves in the graphs on the right column of Fig.~\ref{rescaledMR_caseB}.

\subsection{Model C}

\begin{figure}[h]
 \centering
 \includegraphics[width=8.15cm]{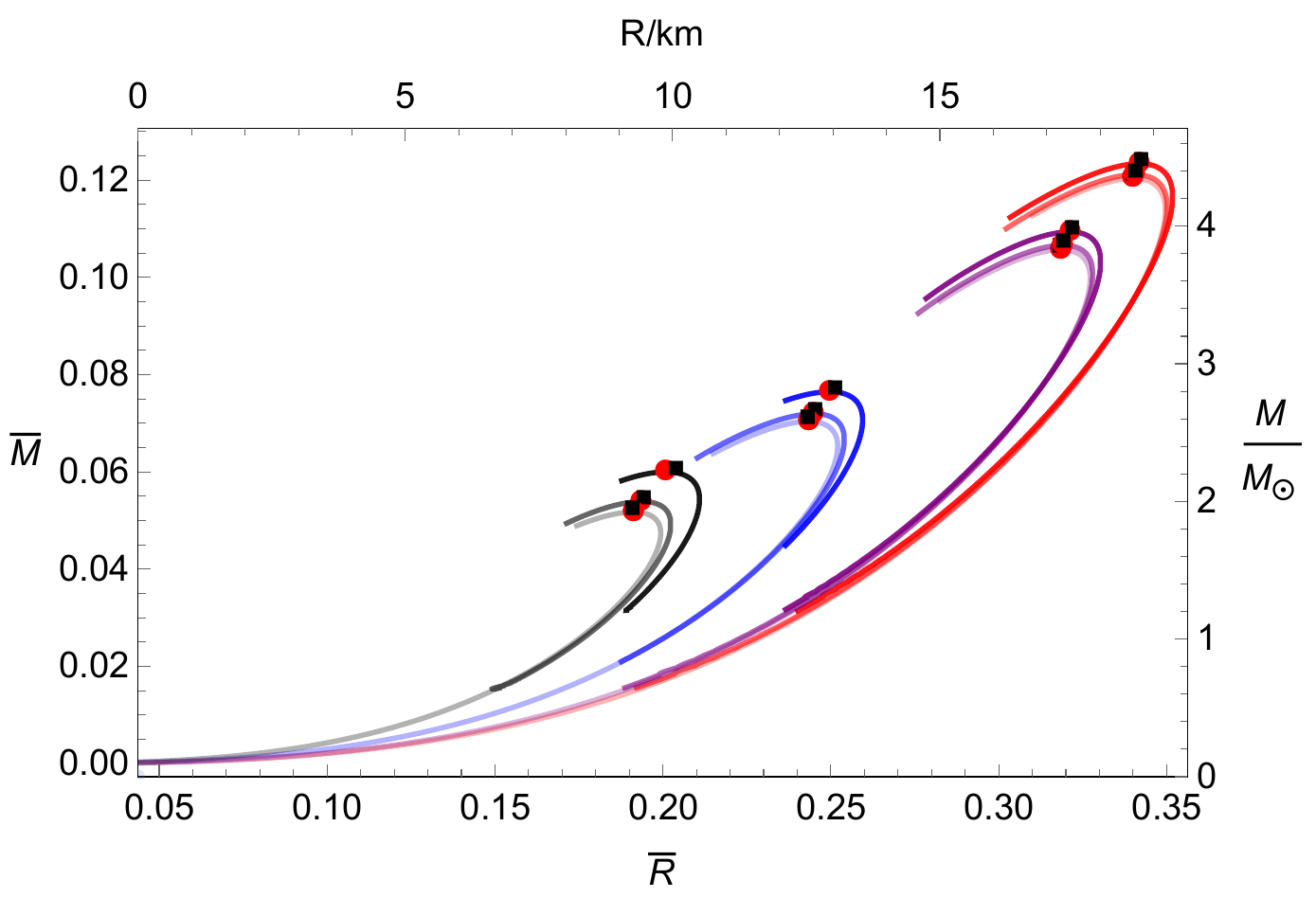}  
  \includegraphics[width=8.15cm]{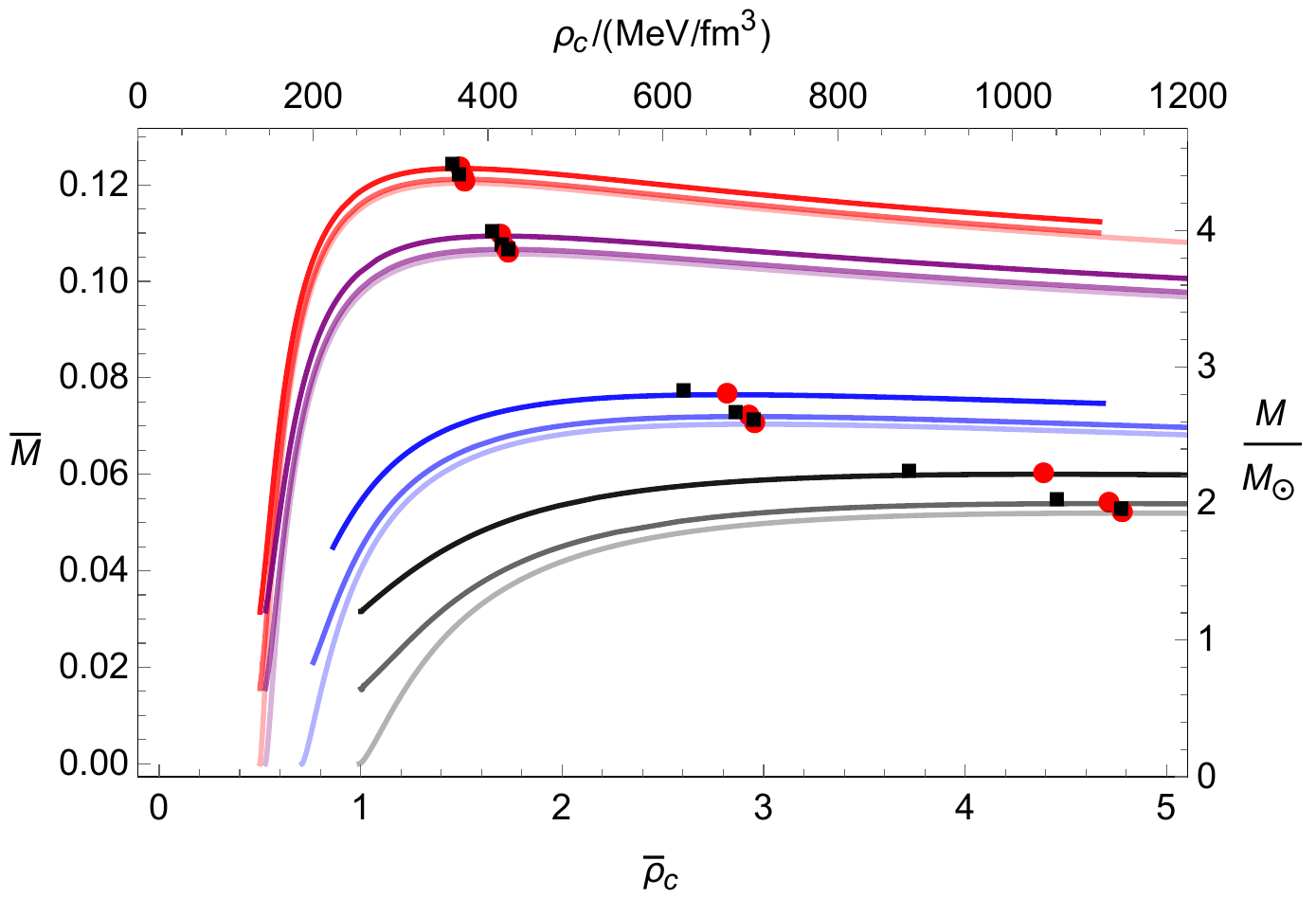}  
\caption[Short caption]{ $\bar{M}$-$\bar{R}$ (left) and $\bar{M}$-$\bar{\rho}_c$ (right) plots of charged interacting quark stars for $\bar{Q}=(0, 1.538,3.076)\times 10^{-2}$. The right and top axes in each plot are the corresponding dimensional parameters with $B_{\rm eff}=60\, \rm MeV/fm^3$ for illustration. The  black, blue, and purple curves respectively denote  $\bar{\lambda}=(0, 0.5 , 10)$ with the sign of $\lambda$ being positive. Red curves are with $\bar{\lambda}=\infty$. Darker shades correspond to increasing values of $\bar{Q}$.
The solid dots denote the maximum mass configurations, with filled squares denoting where $\bar{\omega}^2_0=0$.}
\label{rescaled_caseC}
\end{figure}

In this case we fix the total charge, whose actual values are chosen to be commensurate with the previous two charge configurations.  The corresponding results are obtained by enumerating the charge configurations (like model B\footnote{We find for the same fixed charges, that the results for $(\bar{M},\bar{R})$ derived from model A differ from those of  model B at $0.1\%$-$1\%$ order, with the former having slightly larger maximum masses.}) and central densities that yield the chosen fixed charge value for a given $\bar{\lambda}$.  We illustrate the results  in Fig.~\ref{rescaled_caseC}, where we successively chose  $\bar{Q}=(0, 1.538,3.076)\times 10^{-2}$, corresponding to $Q=(0,1, 2)\times 10^{20} C$ for a typical bag constant value $B_{\rm eff}=\rm 60\, MeV/fm^3$.  

As before, both increasing $\bar{\lambda}$ and increasing charge enlarges the mass and radius of the star, and leads to a stiffer slope for the $\bar{M}$-$\bar{\rho}_c$ curves,  yielding a smaller central density at the maximum mass point. Interestingly, similar to model B, the stellar structure with nearly zero central pressure has non-zero radii and masses (i.e., $\bar{M}$-$\bar{R}$ curves do not start from the origin). This can be understood in the sense that, for a fixed $\bar{Q}$, a small radius requires a very large $\bar{\beta}$ since $\bar{Q}=\bar{\beta} \bar{r}^3$, yet our previous discussion of model B has shown both numerically and analytically that very large $\bar{\beta}$ cases can not have radii close to the origin.

We can see that a larger charge pushes the $\bar{\omega}^2_0=0$ point to smaller central densities than that of the maximum mass point, in contrast to an earlier investigation for charged strange quark stars in the context of the MIT bag model~\cite{Arbanil:2015uoa},
but in accord with a more recent investigation of these objects with an interacting EOS~\cite{Goncalves:2020joq}.
Interestingly, we observe that a larger $\bar{\lambda}$ tend to move the $\bar{\omega}^2_0=0$ point closer to the maximum mass point. This means that for a fixed total charge strong interaction effects tend to offset charge  destabilization effects on radial stability, as expected from the confining nature of the strong interaction. 

\section{results for negative $\lambda$}
\begin{figure}[htb]
 \centering
  \includegraphics[width=8.15cm]{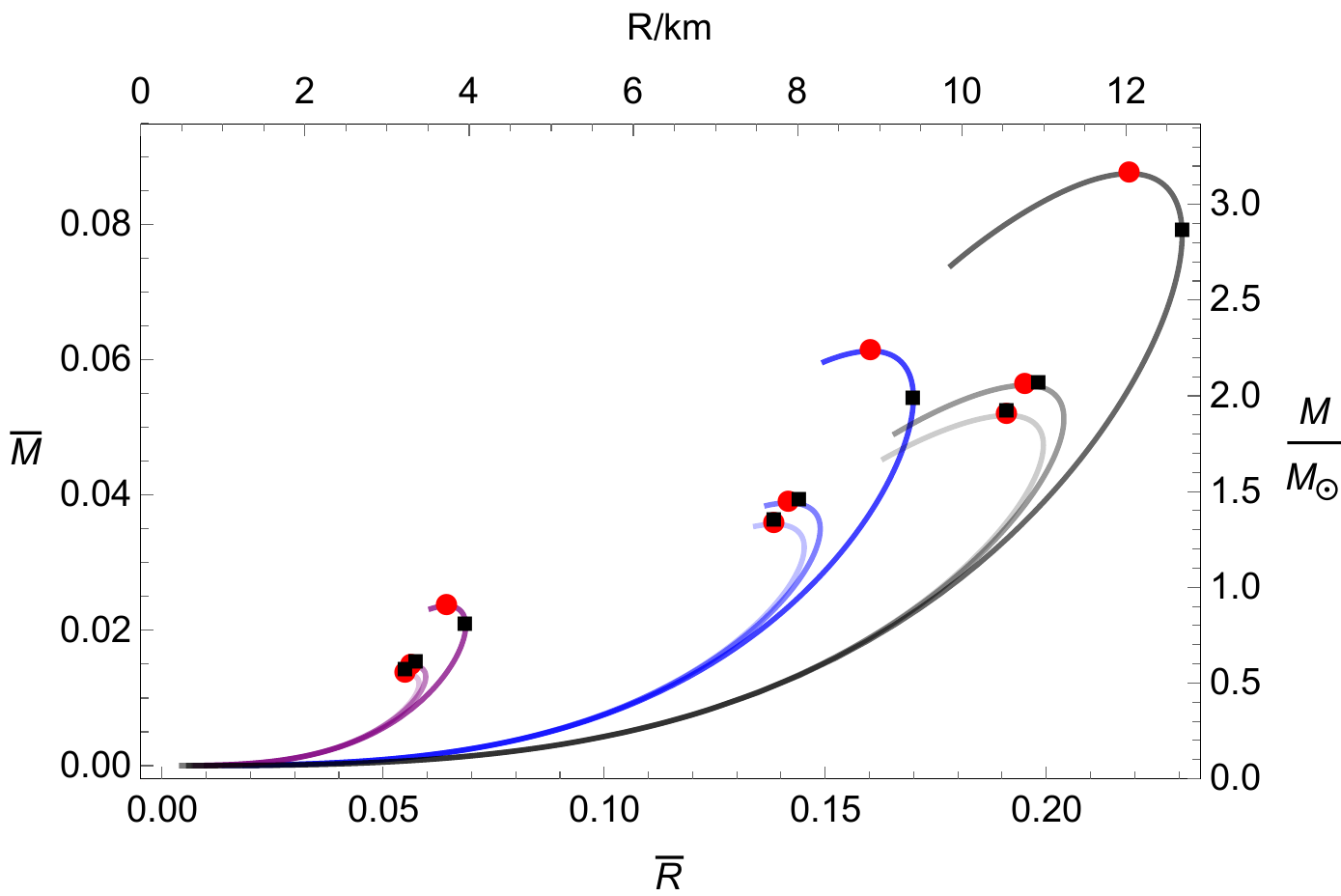}  
    \includegraphics[width=8.15cm]{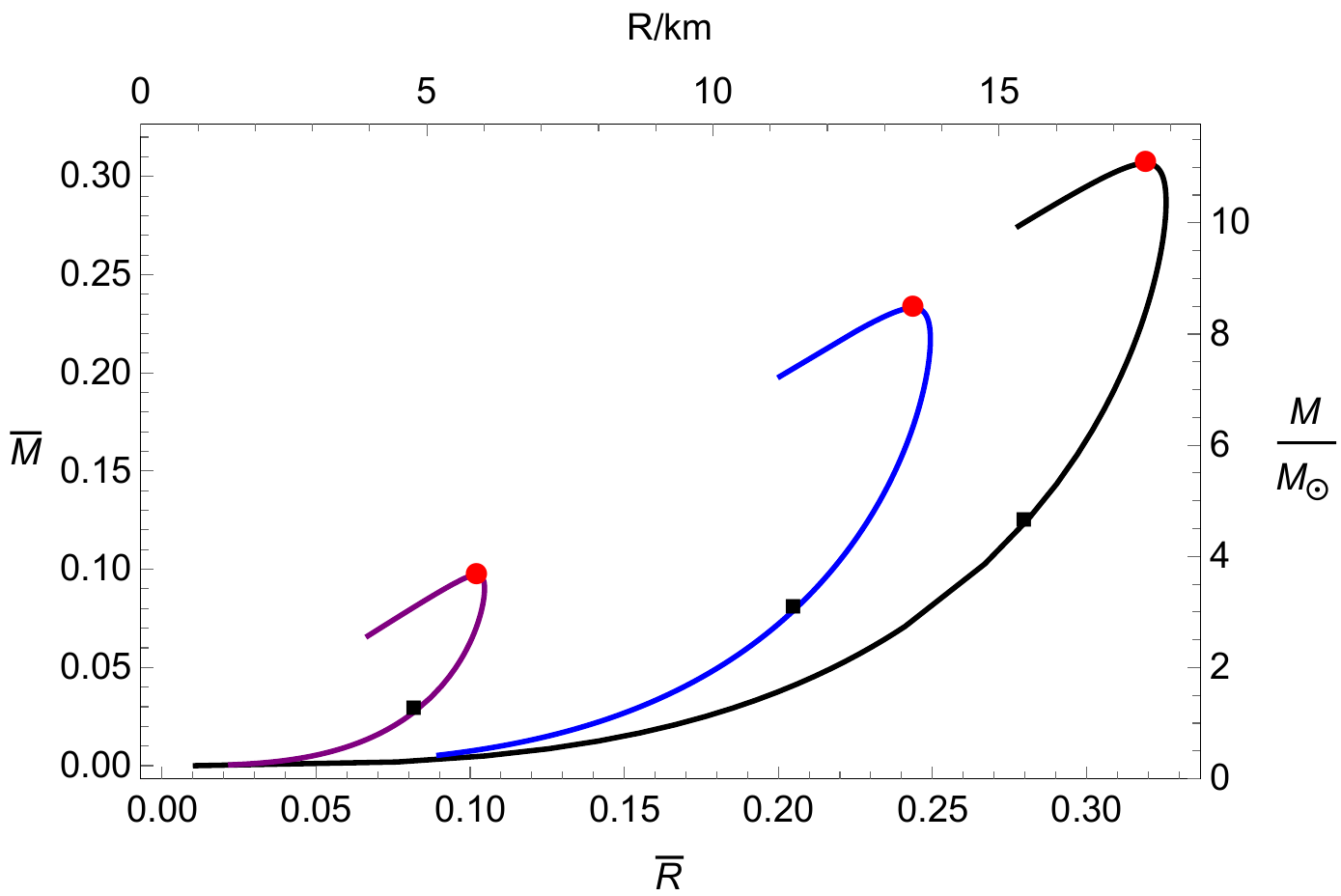}  
   \includegraphics[width=8.15cm]{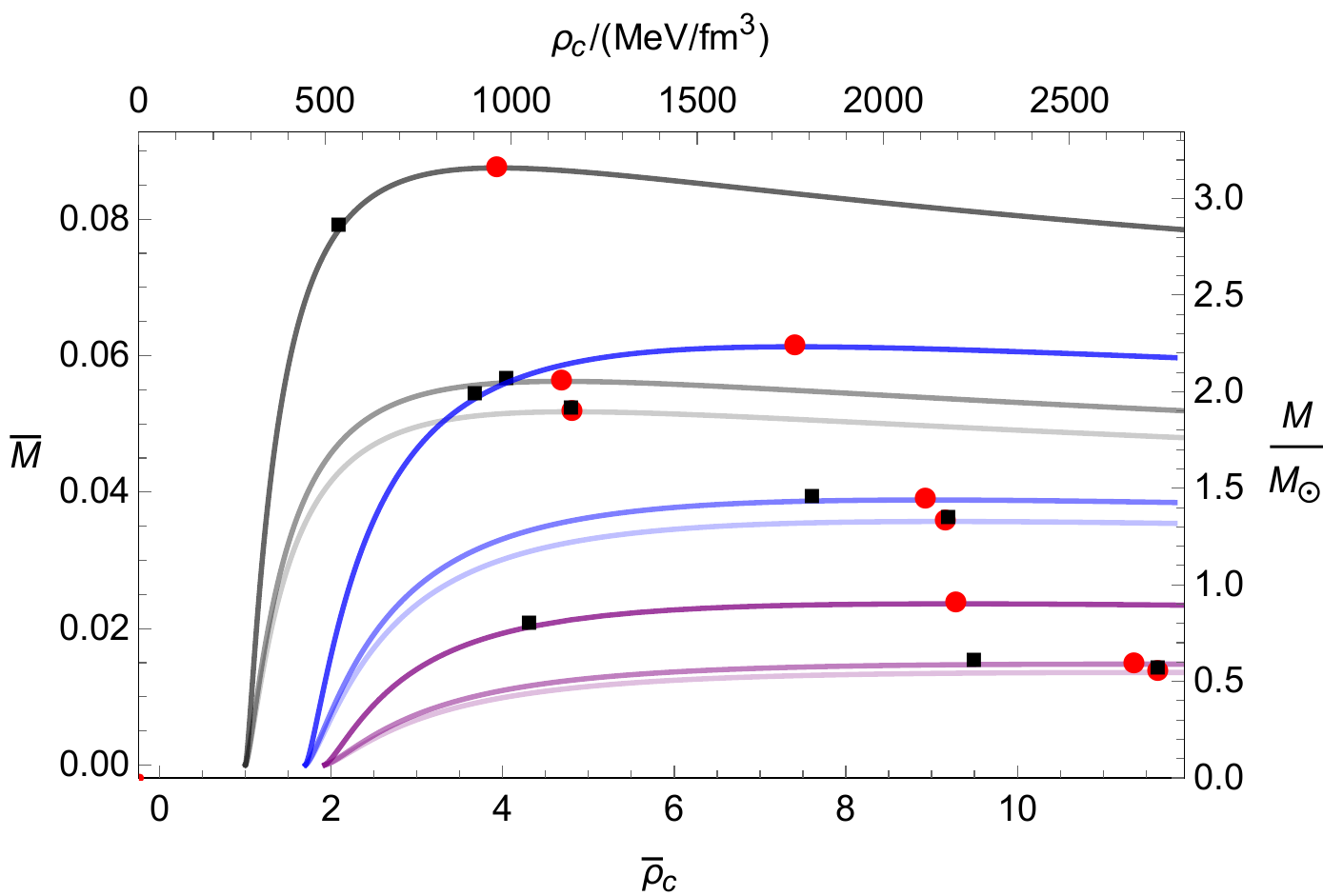}  
      \includegraphics[width=8.15cm]{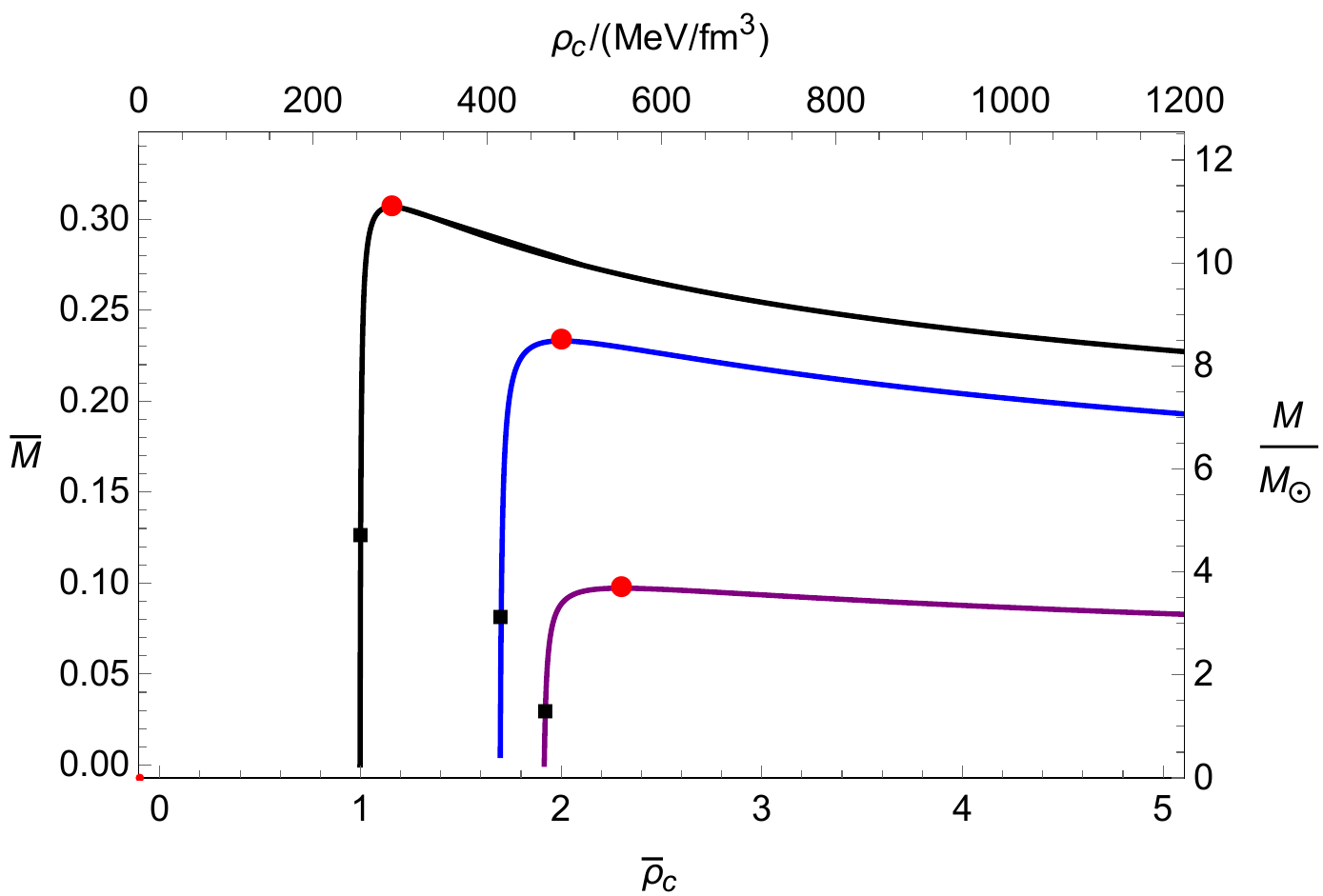}  
\caption{Plots of $\bar{M}$ vs. $\bar{R}$ (upper graphs) and $\bar{\rho}_c$ (lower graphs) of charged interacting quark stars in model A for (left) $\alpha=(0, 0.3, 0.7)$ and (right) $\alpha=0.999$. The right and top axes in each plot are the corresponding dimensional parameters with $B_{\rm eff}=60\, \rm MeV/fm^3$ for illustration. The  black, blue, and purple curves respectively denote $\bar{\lambda}=(0, 0.5 , 10)$ with the sign of $\lambda$ being negative. Darker shades correspond to increasing values of $\alpha$. The solid dots denote the maximum mass configurations, with the filled squares representing where $\bar{\omega}^2_0=0$. Note that for the purple curves in the $\bar{M}-\bar{\rho}_c$ plots, we rescaled the axis as $\bar{\rho}_c\to\bar{\rho}_c/5$ for a clear illustration.}
\label{mA_negLam}
\end{figure}

\begin{figure}[h]
 \centering
 \includegraphics[width=8.15cm]{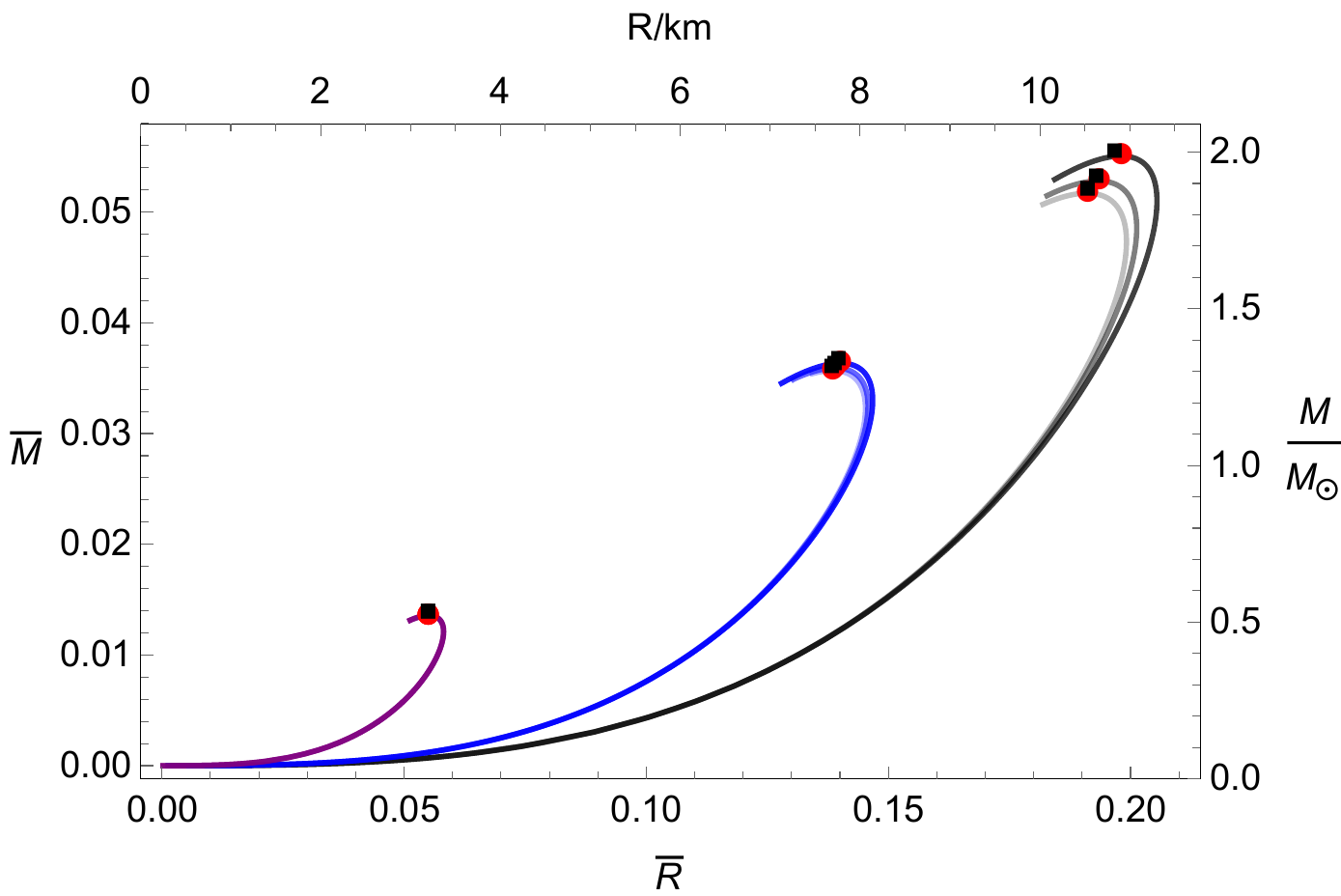}  
     \includegraphics[width=8.15cm]{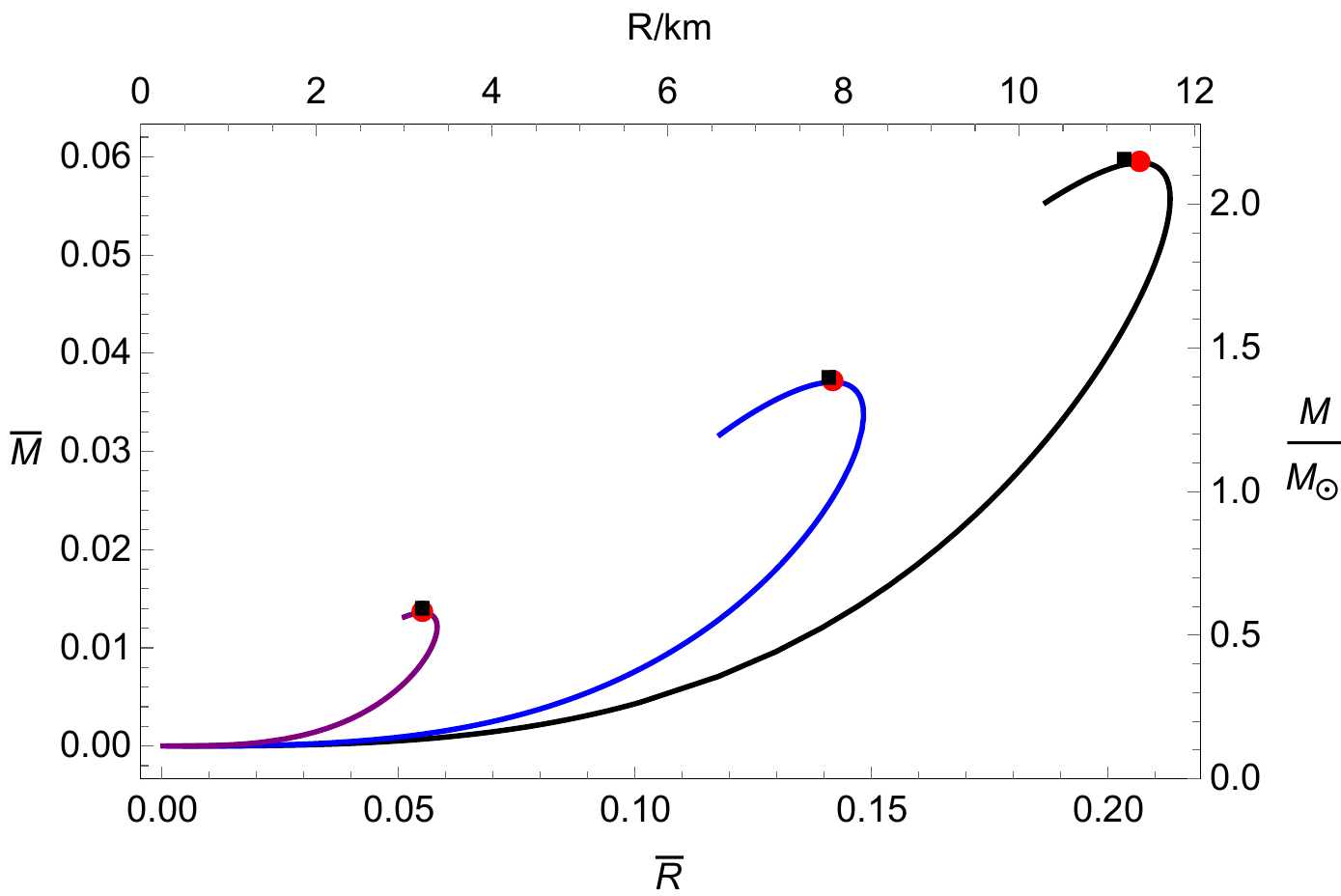}  
   \includegraphics[width=8.15cm]{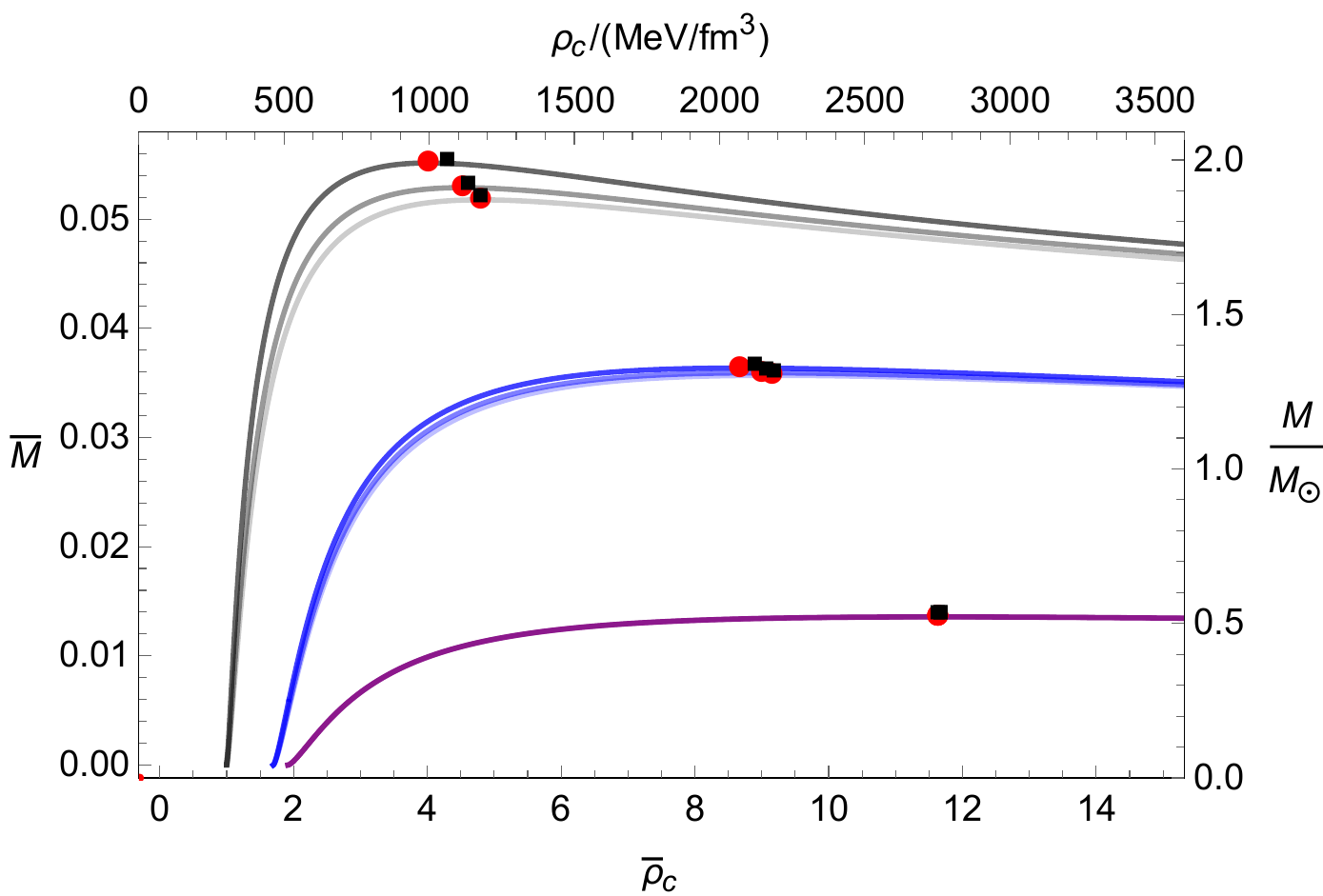}  
     \includegraphics[width=8.15cm]{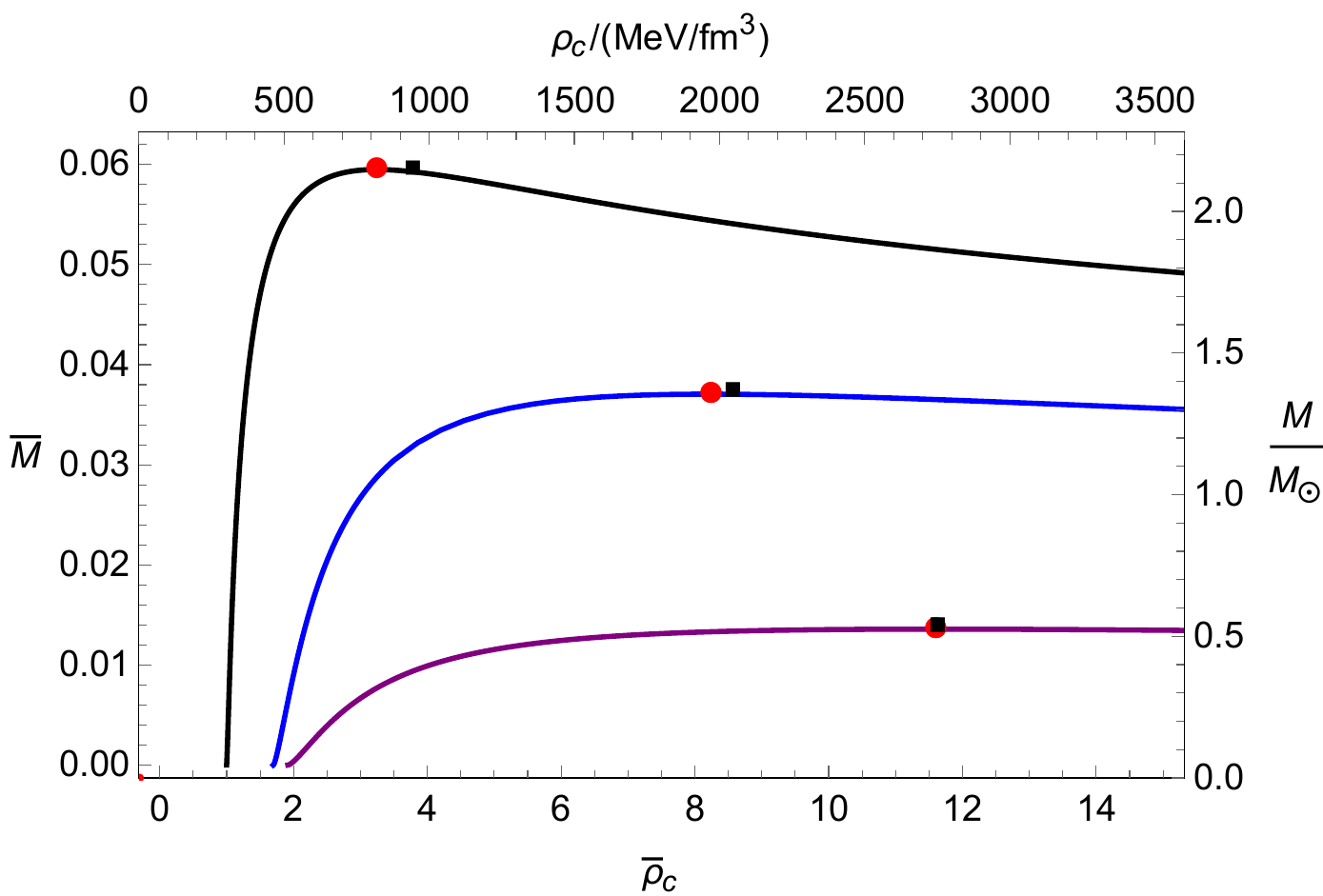}  
\caption{Plots of $\bar{M}$ vs. $\bar{R}$ (upper graphs) and $\bar{\rho}_c$ (lower graphs) of charged interacting quark stars in model B for (left) $\bar{\beta}=(0,1.5, 2.5)$  and (right) $\bar{\beta}=3.5$. The right and top axes in each plot are the corresponding dimensional parameters with $B_{\rm eff}=60\, \rm MeV/fm^3$ for illustration. The black, blue, and purple curves respectively denote $\bar{\lambda}=(0, 0.5 , 10)$ with the sign of $\lambda$ being negative. Darker shades correspond to increasing values of $\bar{\beta}$. 
The solid dots denote the maximum mass configurations, with the filled squares representing where $\bar{\omega}^2_0=0$. Note that for the purple curves in the $\bar{M}-\bar{\rho}_c$ plots, we rescaled the axis as $\bar{\rho}_c\to\bar{\rho}_c/5$ for a clear illustration.}
\label{mB_negLam}
\end{figure}

\begin{figure}[h]
 \centering
   \includegraphics[width=8.15cm]{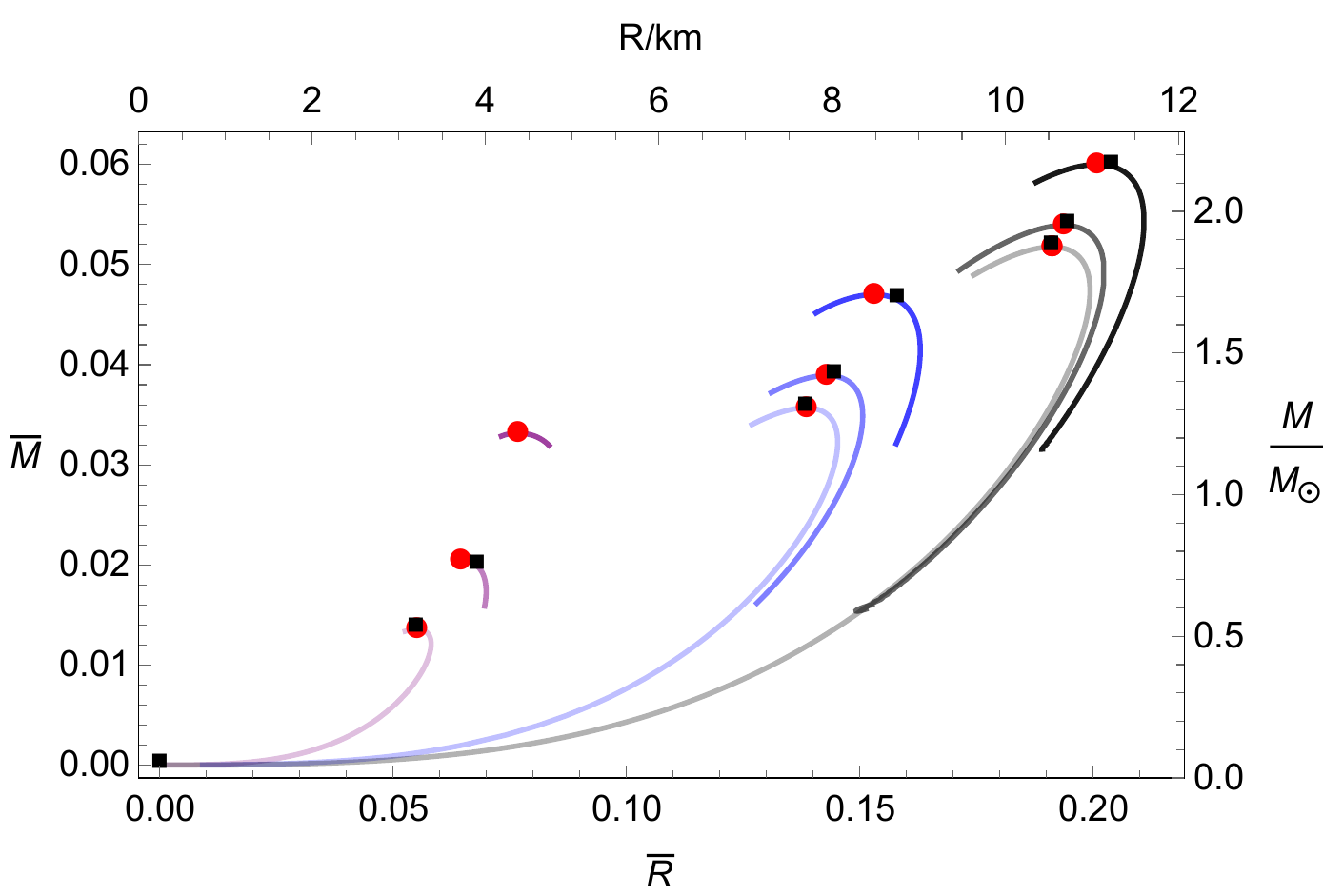}  
  \includegraphics[width=8.15cm]{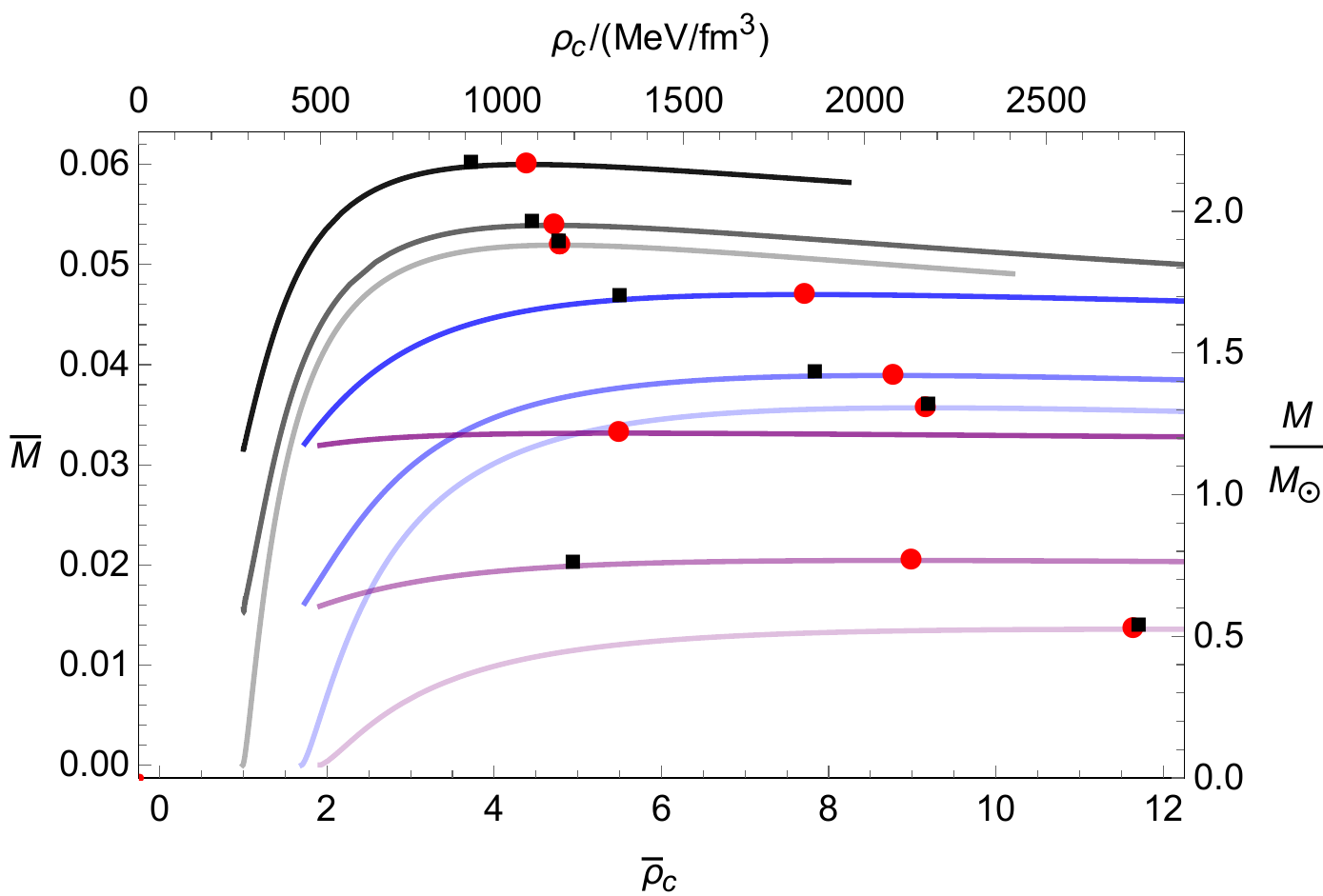}  
\caption{$\bar{M}$-$\bar{R}$ (left) and $\bar{M}$-$\bar{\rho}_c$ (right) of charged interacting quark stars for $\bar{Q}=(0, 1.538,3.076)\times 10^{-2}$. The right and top axes in each plot are the corresponding dimensional parameters with $B_{\rm eff}=60\, \rm MeV/fm^3$ for illustration. The  black, blue, and purple curves respectively denote  $\bar{\lambda}=(0, 0.5 , 10)$ with the sign of $\lambda$ being negative. Darker shades correspond to increasing values of $\bar{Q}$. The solid dots denote the maximum mass configurations, with filled squares denoting where $\bar{\omega}^2_0=0$. Note that for the purple curves in the $\bar{M}-\bar{\rho}_c$ plots, we rescaled the axis as $\bar{\rho}_c\to\bar{\rho}_c/5$ for a clear illustration.}
\label{mC_negLam}
\end{figure}

Here we present the results of the stellar structure and the radial stability for charged interacting quark stars with a negative $\lambda$.  Fig.~\ref{mA_negLam}, Fig.~\ref{mB_negLam}  and Fig.~\ref{mC_negLam} are associated with charge configurations of models A, B, and C, respectively.

First, in all of these figures, we observe that the effect of charge is to lift the mass and radius. However, in contrast to the positive $\lambda$ cases, here a larger $\bar{\lambda}$ results in a smaller mass and radius, and leads to a softer slope for the $\bar{M}$-$\bar{\rho}_c$ curves, resulting in a larger central density at the maximum mass point. This is as expected considering a larger $\bar{\lambda}$ corresponds to a softer EOS when $\lambda$ is negative. We also note that for the model B here, $\bar{M}-\bar{R}$ and $\bar{M}-\bar{\rho}_c$ curves with different $\bar{\beta}$ tend to overlap for a larger $\bar{\lambda}$, which explains why we only observe only one purple curve in Fig.~\ref{mB_negLam}.  

Comparing Fig.~\ref{mB_negLam} to Fig.~\ref{rescaledMR_caseB} for the same model-B-type charge configurations, we can see another obvious difference to the positive $\lambda$ cases is that now all the curves for model B do start from the origin $(\bar{M},\bar{R})=(0,0)$. This is because, when the sign of $\lambda$ turns negative, the sampled $\bar{\beta}$ now ($\leq 3.5$) is not large enough to exceed the critical beta $\bar{\beta}_{c0}$ beyond which things turn otherwise. To be more explicit, restoring the sgn$(\lambda)$ factor in front of the square root term in Eq.~(\ref{betac}), the Eq.~(\ref{betac0}) for the positive $\lambda$ cases then converts to 
\be
\bar{\beta}_{c0}= \frac{4\pi}{3}\left(1- \frac{1}{1-\sqrt{1+\pi^2/(4\bar{\lambda})}}\right)
\label{betac0_neglam}
\ee
for the negative $\lambda$ cases.  Then for $\bar{\lambda}=(0.5, 10)$ we obtain respectively $\bar{\beta}_{c0}\approx (7.11, 40.12)$, which are well above the sampled $\bar{\beta}$ values used in Fig.~(\ref{mB_negLam}).  We also see that now a  larger $\bar{\lambda}$ maps to a larger $\bar{\beta}_{c0}$, in contrast to the positive $\lambda$ cases discussed under Eq.~(\ref{betac0}).

Finally, from the results of radial stability, we see that the relative position of the $\bar{\omega}^2_0=0$ point and the maximum mass point is similar to those in the positive $\lambda$ cases, i.e.,  the $\bar{\omega}^2_0=0$ point can occur at a larger density than that of the maximum mass point for model B, while the opposite  is true for   model A and   model C. And similarly to the positive $\lambda$ cases, a larger charge configuration results in a larger separation size of the two points. In opposition to the positive $\lambda$ case, for model A and model C, a larger $\bar{\lambda}$ shifts the relative position of the $\bar{\omega}^2_0=0$ point to be further apart from the maximum mass point, whereas these two points are closer for model B.   
We find that less negative values of $\lambda$ map to a decreased density separation size of the two points for models A and C, and an enlarged separation size for model B, which decreases  once   $\lambda$  becomes sufficiently large.

\section{Summary}
In this paper, we examined the stellar structure and the radial stability of charged quark stars with different charge configurations and a unified interacting quark matter EOS, which depends on only ($B_{\rm eff}$, $\lambda=(\xi_{2a} \Delta^2-\xi_{2b} m_s^2)/\sqrt{\xi_4 a_4}$) or $\bar{\lambda}=\lambda/4B_{\rm eff}$. In general, a larger charge profile tends to increase the stellar masses and radii. A larger $\bar{\lambda}$ also increases the  mass and radius when $\lambda$ is positive, with the opposite holding for negative $\lambda$. For the charge model with $q=\beta r^3$, we have numerically and analytically identified a new stellar structure with a finite radius at zero center pressure. This exotic structure is more prominent for  larger charge and   larger (smaller) $\bar{\lambda}$ for positive (negative) $\lambda$. Model C (fixed charge $Q$) shows a similar exotic structure since otherwise the results will be inconsistent with those of model B.

Determining the radial stability by identifying the zero eigenfrequencies of the fundamental oscillation mode ($\bar{\omega}^2_0=0$), we observe that stable structures can occur beyond the maximum mass point for model B, while the opposite is true for models A and C, with a larger mass separation (between the maximum mass point and the $\bar{\omega}^2_0=0$ point) size for a larger charge configuration.

 We find that an increasingly positive $\lambda$ tends to decrease the  density separation size of the two points for models A and C, whereas  this separation for model B becomes larger, until  $\lambda$ becomes sufficiently large, after which it decreases.
 This indicates that a large $\lambda$, which maps to a large $\Delta$ or small $m_s$ for a given $a_4$, tends to offset the opposite effects of charge on the star's radial stability. 

\begin{acknowledgments}
We thank Lucas Lazzari for helpful discussions. This work was supported in part by the Natural Sciences and Engineering Research Council of Canada.
\end{acknowledgments}

\end{document}